\documentclass[twocolumn,times]{aastex701}

\usepackage{multirow}
\usepackage{amsmath}
\usepackage{xcolor}

\hypersetup{linkcolor=blue,citecolor=blue}

\def\equationautorefname~#1\null{Eq.~(#1)\null}

\begin{document}
  
\title{The Role of Galactic Shear in Regulating Local Star Formation Efficiency: A Combined MaNGA DynPop and ALMaQUEST Study}

\correspondingauthor{Kai Zhu} 

\author[0000-0002-2583-2669]{Kai Zhu}
\affiliation{Department of Astronomy, School of Science, Westlake University, Hangzhou 310030, Zhejiang Province, People’s Republic of China}
\affiliation{Institute of Natural Sciences, Westlake Institute for Advanced Study, Hangzhou 310024, Zhejiang Province, People’s Republic of China}
\email[show]{zhukai@westlake.edu.cn}
  
\author[0000-0001-8317-2788]{Shude Mao}
\affiliation{Department of Astronomy, School of Science, Westlake University, Hangzhou 310030, Zhejiang Province, People’s Republic of China}
\email{shude.mao@westlake.edu.cn}

\author[0000-0002-1283-8420]{Michele Cappellari}
\affiliation{Sub-department of Astrophysics, Department of Physics, University of Oxford, Denys Wilkinson Building, Keble Road, Oxford, OX1 3RH, UK}
\email{michele.cappellari@physics.ox.ac.uk}

\author[0000-0002-1768-1899]{Sara L. Ellison}
\affiliation{Department of Physics \& Astronomy, University of Victoria, Finnerty Road, Victoria, BC V8P 1A1, Canada}
\email{sarae@uvic.ca}

\author[0009-0004-6271-4321]{Tao Jing}
\affiliation{Department of Astronomy, Tsinghua University, Beijing 100084, People’s Republic of China}
\email{tao.jing.astro@gmail.com}

\author[0000-0001-7218-7407]{Lihwai Lin}
\affiliation{Institute of Astronomy \& Astrophysics, Academia Sinica, No.1, Sec. 4 Roosevelt Road, Taipei 10617, Taiwan}
\email{lihwailin@asiaa.sinica.edu.tw}

\author[0000-0002-6726-9499]{Shengdong Lu}
\affiliation{Institute for Computational Cosmology, Department of Physics, University of Durham, South Road, Durham, DH1 3LE, UK}
\email{shengdong.lu@durham.ac.uk}

\author[0000-0003-0080-8547]{Mallory D. Thorp}
\affiliation{Argelander-Institut f\"ur Astronomie, Universit\"at Bonn, Auf dem H\"ugel 71, 53121 Bonn, Germany}
\email{mthorp@uni-bonn.de}

\author[0000-0002-8614-6275]{Yong Shi}
\affiliation{Department of Astronomy, School of Science, Westlake University, Hangzhou 310030, Zhejiang Province, People’s Republic of China}
\email{shiyong@westlake.edu.cn}

\author[0000-0003-3899-0612]{Ran Li}
\affiliation{School of Physics and Astronomy, Beijing Normal University,  Beijing 100875, People’s Republic of China}
\affiliation{School of Astronomy and Space Science, University of Chinese Academy of Sciences, Beijing 100049, People’s Republic of China}
\email{liran827@icloud.com}

\author[0000-0002-1370-6964]{Hsi-An Pan}
\affiliation{Department of Physics, Tamkang University, No.151, Yingzhuan Road, Tamsui District, New Taipei City 251301, Taiwan}
\email{hapan@gms.tku.edu.tw}

\begin{abstract}  
We study the local regulation of star formation efficiency (SFE) in star-forming regions of nearby galaxies using spatially resolved data from the Mapping Nearby Galaxies at Apache Point Observatory (MaNGA) and the ALMA-MaNGA QUEnching and STar formation (ALMaQUEST) surveys. We define the local SFE as $\Sigma_{\rm SFE} \equiv \Sigma_{\rm SFR}/\Sigma_{\rm H_2}$, where $\Sigma_{\rm SFR}$ is the star formation rate surface density derived from dust-corrected $\rm H\alpha$ emission, and $\Sigma_{\rm H_2}$ is the molecular hydrogen surface density derived from CO measurements. Combining these data with dynamical properties from the MaNGA DynPop project, we use random forest regression to identify the dominant factors of $\Sigma_{\rm SFE}$. Galactic shear, quantified by the Oort parameter $A_{\rm Oort}$, emerges as the dynamical property most strongly associated with lower SFE on average, exceeding the predictive power of mass surface density (either total or stellar), Toomre stability, and mid-plane pressure. However, these quantities provide only modest predictive power and primarily trace the radial variation of SFE. After removing the shear-related radial trends, only weak correlations remain between SFE and other local galactic properties, implying that the remaining azimuthal variations are likely governed by cloud-scale processes below our resolution limit. Furthermore, the relationship between SFE and shear varies among galaxies and weakens in gas-rich regions, suggesting that gas density may modulate the impact of shear. This behavior is consistent with competition between shear-induced cloud stabilization and shear-driven cloud-cloud collisions. Overall, galactic shear appears to regulate the large-scale radial distribution of SFE, while cloud-scale interstellar medium physics may drive the remaining variation.

\end{abstract}


\section{Introduction} \label{sec:intro}
Star formation is one of the key processes in galaxy evolution. It is closely linked to the reservoir of cold gas within and around galaxies, as well as to the physical conditions that govern the conversion of this gas into stars. Among these factors, the availability of cold gas is generally considered the primary determinant of the star formation rate (SFR), as it provides the fuel required for ongoing star formation \citep[e.g.,][]{Schmidt1959,Kennicutt1998,Silk1997,Elmegreen1997,Shi2011,Shi2018}. Star-formation quenching mechanisms therefore act either by reducing the supply of cold gas or by suppressing its conversion into stars. However, the causes of star-formation quenching are highly complex, and whether quenching is dominated by a single mechanism or by a combination of multiple processes remains under debate.

Various mechanisms that suppress star formation may dominate in different types of galaxies and lead to different observational features. Environmental quenching mechanisms such as ram pressure stripping \citep{Gunn1972,Vogt2004,Chung2009} and tidal stripping \citep{Vogt2004,Wetzel2013} are more efficient for satellite galaxies in dense environments. For high-mass galaxies, the dominant quenching mechanisms include, but are not limited to, halo quenching induced by shock heating of infalling gas \citep{Dekel2006,Woo2013} and active galactic nucleus (AGN) feedback \citep{Harrison2017,Lammers2023}. These mechanisms require high-mass dark matter (DM) halos or supermassive black holes (SMBHs), both of which are strongly correlated with galaxy mass. Meanwhile, stellar feedback driven by stellar winds, supernovae, or radiation pressure \citep{Colling2018,Ostriker2022} is usually thought to be more efficient in low-mass galaxies with shallow gravitational potential wells. The presence of massive dense structures at galactic centers may also play a major role in halting star formation, such as bar quenching \citep{Khoperskov2018} and morphological quenching \citep{Martig2009,Martig2013,Gensior2020,Gensior2021} that occurs when the gas disk is stabilized by the growth of a stellar spheroid or bulge \citep{Toomre1964}. Morphological quenching, also known as gravitational quenching \citep{Genzel2014, Kalinova2022}, does not involve the removal or heating of cold gas and may explain observations of quenched galaxies that still retain a large amount of cold gas \citep[e.g.,][]{Serra2012,Young2014,Wang2022,Sharma2023,LiXiao2024,Dou2024,Jing2024,LiYang2025}, i.e., galaxies with a low star formation efficiency ($\rm SFE\equiv SFR/M_{\rm H_2}$). 

Galactic shear, which measures the relative tangential velocity between two points at different galactocentric radii, is also expected to suppress the SFR by reducing the SFE. Galactic shear acts on the scale of giant molecular clouds (GMCs) by stabilizing their self-gravity and preventing their gravitational collapse \citep[e.g.,][]{Davis2014,Colombo2018,Colling2018,Liu2021,LuAnan2024,LuAnan2025,Ni2025}, by disrupting GMCs before they collapse and form stars \citep{Hunter1998,Chevance2023}, or by driving turbulence that counterbalances the self-gravity of GMCs \citep{Zhou2017,Gensior2023,Rani2024}. However, several studies disfavor the scenario that shear universally suppresses star formation. For example, \citet{Tan2000} proposed a shear-driven cloud-cloud collision model, in which star formation occurs in regions of compressed gas during collisions between GMCs. The cloud-cloud collision model predicts a different correlation between SFR and shear, suggesting that higher shear leads to higher SFR at a fixed gas surface density. Such a positive influence of shear on star formation has been reported in a number of extragalactic studies \citep[e.g.,][]{Tan2010,Suwannajak2014,Aouad2020,Sun2022,Fortune-Bashee2024,Horie2024}. Moreover, \citet{Dib2012}, in their study of individual molecular clouds in the Milky Way, found no evidence for a correlation between shear and SFR, suggesting that shear plays only a minor role in regulating star formation. Taken together, these conflicting findings indicate that the role of shear in regulating SFE is still controversial and poorly understood.

In this paper, we aim to investigate how galactic shear regulates the local surface density of SFE ($\Sigma_{\rm SFE}$), which is defined as the ratio of the local surface density of SFR ($\Sigma_{\rm SFR}$) to the local surface density of molecular gas ($\Sigma_{\rm H_2}$), based on the integral field unit (IFU) observations from the Mapping Nearby Galaxies at Apache Point Observatory (MaNGA) survey \citep{Bundy2015}. The paper is organized as follows. In \autoref{sec:data}, we describe the data, sample selection, and the calculation of spatially resolved properties. The results are presented in \autoref{sec:results}, including two-dimensional (2D) maps of spatially resolved properties (\autoref{subsec:2Dmaps}), results from random forest regression (\autoref{subsec:RF}), the galaxy-to-galaxy variation in how shear regulates SFE (\autoref{subsec:gal2gal_variation}), and the azimuthal variations of SFE (\autoref{subsec:azimuthal_SFE_variation}). The main conclusions are summarized in \autoref{sec:conlusions}. Throughout this paper, we adopt a flat Universe with $\Omega_{\rm m} = 0.307$ and $H_0 = 67.7\,\mathrm{km~s^{-1}~Mpc^{-1}}$ \citep{Planck2016}, consistent with the MaNGA DynPop series \citep{Zhu2023,Lu2023}. In this paper, $\log_{10}$ is denoted by $\lg$ following the official ISO 80000-2:2019 notation.

\section{Sample and Data} \label{sec:data}
\subsection{The MaNGA survey: MaNGA DynPop and ALMaQUEST}\label{subsec:manga}
The MaNGA survey \citep{Bundy2015} provides IFU observations of $\sim$10,000 nearby galaxies in the final data release of SDSS-IV \citep{SDSSDR17}. The MaNGA spectra cover a wavelength range of $3600-10300\,\mathrm{\AA}$, with a spectral resolution of $\sigma=72\,{\rm km\,s^{-1}}$ \citep{Law2016}. The raw data are reduced with the Data Reduction Pipeline \citep[DRP;][]{Law2016} to produce data cubes with a spaxel size of $0\farcs5\times 0\farcs5$. Additionally, the Data Analysis Pipeline \citep[DAP;][]{Belfiore2019,Westfall2019} applies Voronoi binning \citep{Cappellari2003} to the DRP data cubes to achieve a signal-to-noise ratio of $\rm S/N=10$ and performs full spectral fitting to the Voronoi-binned spectra using \textsc{ppxf} \citep{Cappellari2017}, deriving measurements of stellar kinematics, emission-line properties, and spectral indices.

Using the stellar kinematics derived from the DAP and spatially resolved spectra provided by the DRP, the MaNGA DynPop project \citep{Zhu2023,Lu2023} derived dynamical and stellar population properties for the full sample of MaNGA galaxies. Specifically, \citet{Zhu2023} applied axisymmetric Jeans Anisotropic Modeling \citep[JAM;][]{Cappellari2008,Cappellari2020}, which relates the DAP 2D maps of root-mean-square velocity $V_{\rm rms}^2 $($\equiv V_{\rm los}^2+\sigma_{\rm los}^2$, where $V_{\rm los}$ and $\sigma_{\rm los}$ are the line-of-sight stellar velocity and stellar velocity dispersion) to the gravitational potential, yielding mass distributions of multiple components (including the total, stellar, and DM components) under different assumptions about the DM distribution. In this paper, we adopt the most flexible model with a generalized Navarro-Frenk-White \citep[gNFW;][]{Wyithe2001} DM profile, and we refer readers to \citet{Zhu2023} for more details. By comparing the observed and modeled stellar kinematics, the entire sample is classified into different modeling quality classes (Qual = $-1$, 0, 1, 2, 3, from worst to best). Qual = $-1$ corresponds to highly disturbed systems, Qual = 0 to poor fits caused by data or modelling issues, Qual = 1–2 to progressively better recovery of the kinematics (especially $V_{\rm rms}$), and Qual = 3 to models that successfully reproduce both $V_{\rm rms}$ and $V_{\rm los}$ \citep[see more details in Section~5.1 of][]{Zhu2023}. Among the MaNGA sample, 6065 galaxies with Qual $\geqslant 1$ have total mass distributions that are virtually insensitive to different model assumptions \citep{Zhu2023}. With the JAM-derived gravitational potential, one can calculate the circular velocity curves (i.e. intrinsic rotation curves) or total mass surface density maps, following \citet{Zhu2025}\footnote{The database and a Python script to calculate different types of mass distributions are publicly available in \url{https://manga-dynpop.github.io}}. Moreover, \citet{Lu2023} derived stellar population properties from the MaNGA DRP IFU spectra, using the \textsc{ppxf} software \citep{Cappellari2004,Cappellari2017,Cappellari2023} with the \textsc{fsps} models \citep{Conroy2009,Conroy2010}, the Salpeter \citep{Salpeter1955} initial mass function (IMF), and the MIST isochrones \citep{Choi2016}.

Other ancillary data come from the ALMA-MaNGA QUEnching and STar formation (ALMaQUEST) program \citep{Lin2020}, which provides spatially resolved CO measurements obtained with the Atacama Large Millimeter Array (ALMA) for 46 galaxies selected from the MaNGA survey. The extended ALMaQUEST sample \citep{Ellison2024} also includes 20 additional galaxies observed during ALMA Cycle 7, primarily merging systems \citep{Thorp2022}. Since these galaxies cannot be reliably modeled with stellar dynamical methods, they are not included in our analysis. Among the ALMaQUEST sample, we select galaxies with JAM modeling quality Qual $\geqslant 1$ and further remove one galaxy (plateifu: 8083-9101) with poor constraints on stellar population properties, resulting in a final sample of 29 galaxies. Our sample, defined by Qual $\geqslant 1$, excludes galaxies with highly disturbed kinematics (Qual = $-1$), such as merging and irregular galaxies, as well as those with somewhat regular kinematics that cannot be well modeled by JAM due to inferior data quality or problematic models (Qual = 0). The distributions of basic parameters (SDSS $r$-band magnitude, $u-g$ color, stellar mass, and stellar angular momentum) in the Qual $\geqslant 1$ sample remain similar to those of the full MaNGA sample \citep[][Figure~8]{Zhu2023}, indicating that no obvious systematic bias is introduced.

\begin{figure*}
    \centering
    \includegraphics[width=\textwidth]{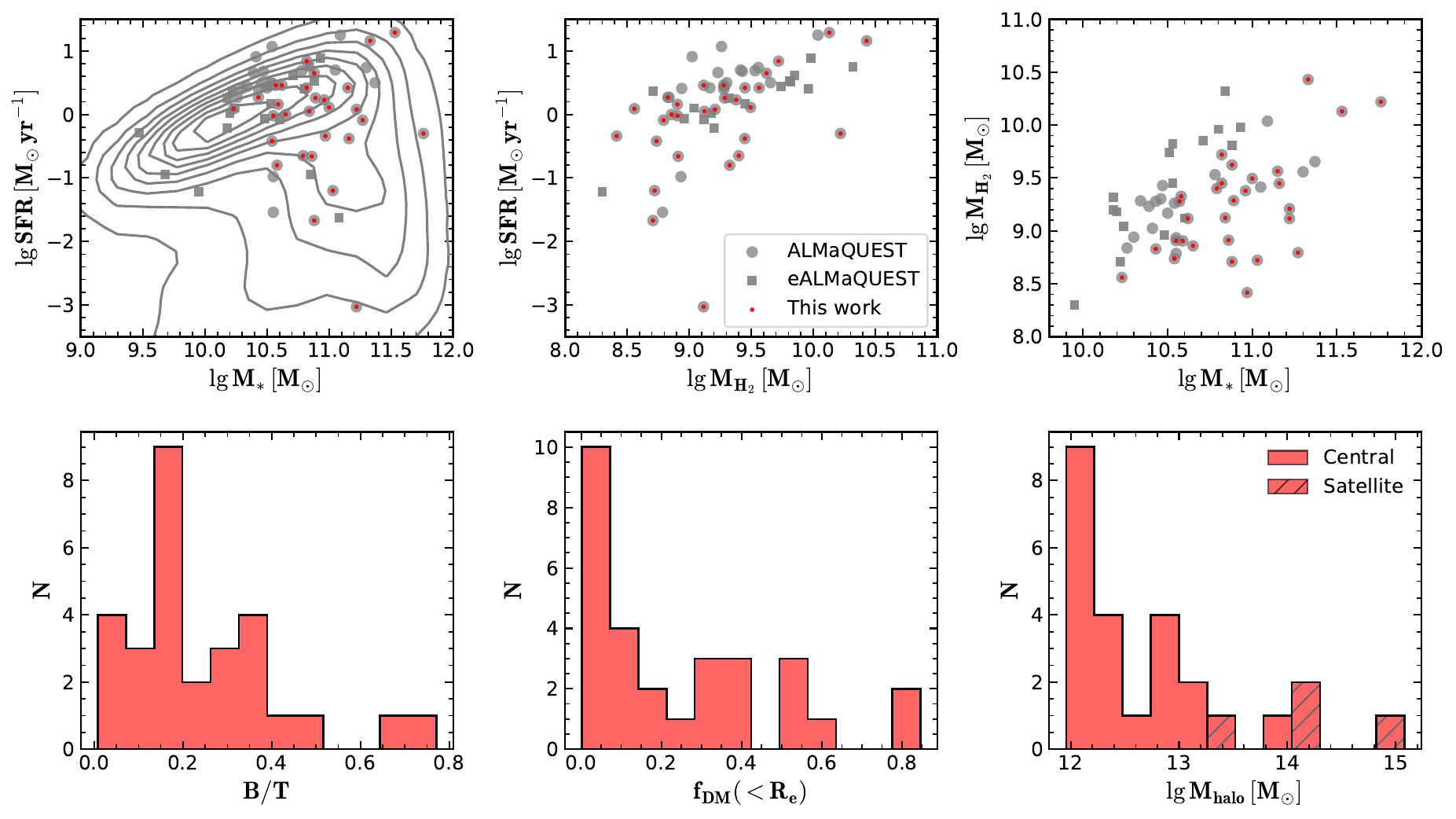}
    \caption{\textit{Top panels:} The scaling relations for the ALMaQUEST sample (gray circles), the extended ALMaQUEST sample (gray squares), and the selected sample in this work (red circles). From left to bottom, the star-forming main sequence (SFR versus $M_{\ast}$), the Kennicutt-Schmidt relation (SFR versus $M_{\rm H_2}$), and the molecular gas main sequence ($M_{\rm H_2}$ versus $M_{\ast}$) are presented, respectively. \textit{Bottom panels:} Histograms of the galactic properties for the selected sample, including the bulge-to-total ratios (left), the dark matter fractions within one effective radius (middle), and the DM halo masses (right). The hatched histogram in the bottom right panel corresponds to the four satellite galaxies.}
    \label{fig:galprop}
\end{figure*}

The top panels of \autoref{fig:galprop} show the global scaling relations for the ALMaQUEST sample (gray circles), the extended ALMaQUEST sample (gray squares), and the selected sample (red circles): the star-forming main sequence \citep[SFMS;][]{Daddi2007,Noeske2007,Whitaker2012}, the Kennicutt-Schmidt (KS) relation \citep{Schmidt1959,Kennicutt1998}, and the molecular gas main sequence \citep[MGMS;][]{Lin2019,Baker2023}. All quantities are computed by summing over the areas within the field-of-view (FoV) of MaNGA, using measurements from \citet[][Table~3]{Lin2020} and \citet[][Table~1]{Ellison2024}. The selected sample spans a similar stellar mass range to the parent ALMaQUEST sample, although its galaxies exhibit slightly lower SFRs and molecular gas masses than the non-selected galaxies, while still remaining on the SFMS. This difference is likely related to the more disturbed stellar kinematics commonly found in gas-rich and starburst systems. 

By matching to the PyMorph photometric catalog \citep{Dominguez-Sanchez2022}, the selected galaxies tend to be disk-dominated, with a median bulge-to-total ratio ($B/T$) of 0.186 and 86$\%$ (25/29) of galaxies having $B/T<0.4$ (bottom left panel of \autoref{fig:galprop}). Although the dark matter fractions within one effective radius $f_{\rm DM}(<R_{\rm e})$ are generally low in MaNGA galaxies \citep{Zhu2024,Lu2024}, our sample still includes a substantial number of galaxies with non-negligible dark matter fractions (12 out of 29 galaxies have $f_{\rm DM}(<R_{\rm e})>30\%$, as shown in the bottom middle panel of \autoref{fig:galprop}). This indicates that the gravitational potential is not solely dominated by the stellar component. We also identify central/satellite galaxies based on the group catalog derived by \citet{Yang2007}, which assigns each galaxy to a group using an adaptive halo-based group finder. Within this sample, there are 24 central galaxies (defined as the most massive galaxy in each group) with a relatively narrow halo mass range of $10^{12}-10^{13}\,{\rm M_{\odot}}$, 4 satellite galaxies, and one galaxy is not assigned to any group (bottom right panel of \autoref{fig:galprop}). These results suggest that environmental effects (e.g., ram pressure stripping, tidal stripping) are unlikely to be the dominant factors regulating star formation in these systems. Overall, the sample is dominated by star-forming, disk galaxies and displays limited environmental diversity.

\subsection{Spatially resolved properties on kpc scales}\label{subsec:spatial_prop}
In the MaNGA DAP, the spatially resolved properties are measured in Voronoi bins (with $\rm S/N = 10$) instead of individual spaxels to obtain reliable measurements of stellar kinematics and emission-line fluxes. We use the total flux measurements of different emission lines ($\rm H\alpha$, $\rm H\beta$, $\rm [O\,\textsc{iii}]$, and $\rm [N\,\textsc{ii}]$) derived from the DAP products, and require the $\rm S/N$ in strong lines ($\rm H\alpha$ and $\rm H\beta$) and weak lines ($\rm [O\,\textsc{iii}]$ and $\rm [N\,\textsc{ii}]$) to be $>3$ and $>2$, respectively. The star-forming (SF), composite, and AGN regions are classified using the line ratios of $\rm [N\,\textsc{ii}]/H\alpha$ and $\rm [O\,\textsc{iii}]/H\beta$ on the Baldwin-Phillips-Terlevich (BPT) diagram \citep{BPT1981}, following the criteria of \citet{Kauffmann2003} and \citet{Kewley2006}. The following analysis focuses on the SF regions. 

The 2D maps of stellar population properties are based on different Voronoi binning performed with a target $\rm S/N$ of 30 \citep{Lu2023}. The stellar population properties obtained for each Voronoi bin are then assigned to all spaxels belonging to that bin, producing spaxel-level maps with values assigned from the corresponding $\rm S/N = 30$ bins. To ensure consistency with the DAP-derived quantities, the stellar population maps are matched to the DAP binning by averaging the stellar population properties over all spaxels associated with each DAP Voronoi bin. Similarly, the spaxel-level CO measurements \citep{Lin2020,Ellison2024} are area-averaged over the DAP Voronoi bins. We additionally require the CO lines to have $\rm S/N > 2$, resulting in 7925 star-forming Voronoi bins that satisfy all of the above $\rm S/N$ criteria. Adopting a more conservative threshold of $\rm S/N > 5$ for the CO lines does not affect our results. The dynamical properties are subsequently computed at the centroids of these Voronoi bins.

For each Voronoi bin, the surface densities (in units of $\rm kpc^{-2}$) of various quantities (e.g., stellar mass, SFR, molecular hydrogen) are corrected for the inclination effects using
\begin{equation}\label{eq:surface_density}
\begin{aligned}
\Sigma_{\rm face-on}
&= \Sigma_{\rm obs}\cos i \\
&= \frac{X\,\cos i}
       {N_{\rm spaxel}
       \left(D_{\rm A}\theta_{\rm spaxel}
       \frac{\pi}{648000}\times10^3\right)^2},
\end{aligned}
\end{equation}
where $X$ is the integrated property measured within the bin, $N_{\rm spaxel}$ is the number of spaxels of this bin, $\theta_{\rm spaxel}$ is the angular size of MaNGA spaxels (0.5 arcsec), $D_{\rm A}$ is the angular diameter distance (in units of Mpc), $\pi/648000$ is the conversion factor from arcseconds to radians, $i$ is the inclination angle ($i=0$ means face-on) determined from JAM models \citep{Zhu2023}. The face-on ($i=0$) surface density of dynamical properties such as the total mass surface density can be derived analytically from the JAM model. The 2D maps of spatially resolved properties used in this paper are summarized below. 

\begin{itemize}
    \item SFR surface density $\Sigma_{\rm SFR}$: The SFR is estimated based on dust-corrected $\rm H\alpha$ flux for those SF Voronoi bins classified by the BPT diagram. Following the method described in \citet[Section~3]{Dominguez2013}, the $\rm H\alpha$ flux (from the MaNGA DAP output) is corrected for dust extinction using the Balmer decrement ($\rm H\alpha/H\beta$). Assuming an intrinsic ratio of $\rm (H\alpha/H\beta)_{\rm intr}=2.86$ \citep[Case B recombination at an electron temperature $T_{\rm e}=10^{4}\,{\rm K}$ and electron density $n_{\rm e}=10^{2}\,{\rm cm^{-3}}$;][]{Osterbrock1989} and adopting the attenuation curve of \citet{Calzetti2000}, the color excess $E(B-V)$ \citep[][Eq.~4]{Dominguez2013} is given by
    \begin{equation}
        E(B-V)=1.97\lg\left[\frac{{\rm (H\alpha/H\beta)_{obs}}}{{\rm (H\alpha/H\beta)_{intr}}}\right],
    \end{equation}
    and the attenuation at $\rm H\alpha$ \citep[][Eq.~6]{Dominguez2013} is
    \begin{equation}
        A_{\rm H\alpha}=3.33\times E(B-V).
    \end{equation}
    The extinction-corrected $\rm H\alpha$ luminosity is
    \begin{equation}
        L_{\rm H\alpha, intr} = L_{\rm H\alpha, obs}10^{0.4A_{\rm H\alpha}},
    \end{equation}
    and the SFR (assuming a Salpeter IMF) is estimated using the conversion factor of \citet{Kennicutt1998ARAA}, written as
    \begin{equation}
        {\rm SFR\,[M_{\odot}\,yr^{-1}]} = 7.9\times10^{-42}L_{\rm H\alpha, intr}[{\rm erg\,s^{-1}}].
    \end{equation}

    \item Gas-phase metallicity $\rm 12 + \lg(O/H)$: The gas-phase metallicity is estimated from the O3N2 index \citep[][Eqs.~1 and 2]{Marino2013}, with 
    \begin{equation}
    \rm O3N2=\lg\left(\frac{[O\,\textsc{iii}]}{H\beta}\times \frac{H\alpha}{[N\,\textsc{ii}]}\right)
    \end{equation}
    and
    \begin{equation}
    \rm 12+\lg(O/H)=8.533[\pm0.012]-0.214[\pm0.012]\times O3N2.
    \end{equation}
    The gas-phase metallicity measurements are not involved in the following analysis, and are only used to derive the metallicity-dependent CO-to-$\rm H_{2}$ conversion factor and the metallicity-dependent atomic-to-molecular gas ratio, as described below. Adopting a more stringent $\rm S/N$ threshold of 5 for $\rm H\alpha$, $\rm H\beta$, $\rm [O,\textsc{III}]$, and $\rm [N,\textsc{II}]$ emission lines does not alter our conclusions.
    
    \item Surface mass density of molecular hydrogen $\Sigma_{\rm H_2}$: The CO(1-0) intensities \citep{Lin2020} are converted into molecular hydrogen ($\rm H_{2}$) surface densities adopting a CO-to-$\rm H_2$ conversion factor of $\rm \alpha_{CO}=4.3\,M_{\odot}\,(K\,km\,s^{-1}\,pc^{2})^{-1}$ \citep{Bolatto2013}. We also test a gas-phase metallicity-dependent CO-to-$\rm H_2$ conversion factor of $\alpha_{{\rm CO}, Z}=4.35\,Z^{-1.6}\,{\rm M_{\odot}\,(K\,km\,s^{-1}\,pc^{2})^{-1}}$ \citep[][Eq.~2]{Sun2020a}, where $Z$ is the O3N2 calibration of gas-phase metallicity normalized by a solar value of $\rm 12+\lg(O/H)=8.69$. The $\Sigma_{\rm H_2}$ values obtained under the two assumptions for the CO-to-$\rm H_2$ conversion factor are in good agreement, with a best-fit slope of 0.94 and a 1$\sigma$ error of 0.043 dex derived using the \textsc{lts\_linefit} procedure \citep{Cappellari2013a}. Consequently, the adopted conversion factor does not significantly affect our results or conclusions.

    \item Surface mass density of cold gas $\Sigma_{\rm gas}$: $\Sigma_{\rm gas}=\Sigma_{\rm H_2}+\Sigma_{\rm HI}$, where $\Sigma_{\rm HI}$ is the atomic hydrogen gas surface density. In the absence of spatially resolved $\Sigma_{\rm HI}$ measurements, we adopt a constant value of $\Sigma_{\rm HI}=7\times10^6\,{\rm M_{\odot}\,kpc^{-2}}$ for all Voronoi bins following \citet{Ellison2024}, which is typical for nearby disk galaxies \citep[e.g.,][]{Bigiel2008}. We also test a metallicity-dependent atomic-to-molecular gas ratio \citep{Yu2024} to derive a spatially varying $\Sigma_{\rm HI}$ from $\Sigma_{\rm H_2}$, and find that this alternative prescription does not affect our conclusions. This quantity is used in subsequent calculations of the \citet{Toomre1964} parameter and mid-plane dynamical equilibrium pressure (see the definitions below).
    
    \item Line-of-sight velocity dispersion of molecular hydrogen $\sigma_{\rm H_2}$: The velocity dispersion of $\rm H_{2}$ is assumed to be equal to that measured from the velocity dispersion (moment-2) maps of the CO(1-0) data cubes \citep{Lin2020}.
    
    \item Spatially resolved SFE $\Sigma_{\rm SFE}$: The local SFE is defined as $\Sigma_{\rm SFE}\equiv \Sigma_{\rm SFR}/\Sigma_{\rm H_2}$. The spatial scales of these measurements are matched to the DAP Voronoi bin, as described earlier in this section. The $\Sigma_{\rm SFE}$ values derived using either a constant or a metallicity-dependent CO-to-$\rm H_2$ conversion factor are highly consistent, exhibiting a best-fit slope of 0.99 and a 1$\sigma$ error of 0.046 dex.
    
    \item Stellar mass surface density $\Sigma_{\ast}$: The spaxel-level stellar mass maps are derived from stellar population synthesis (SPS) model for the $\rm S/N=30$ Voronoi bins and then assigned to individual spaxels within each bin \citep[][keyword: \texttt{Mstar\_Map} in Table~B1]{Lu2023}, assuming a Salpeter IMF. We match these maps to the DAP $\rm S/N=10$ Voronoi binning (see the details described above) and then compute the stellar mass surface density using \autoref{eq:surface_density}. We verify that the resulting matched stellar population maps are nearly identical to the original ones.
    
    \item Total mass surface density $\Sigma_{\rm total}$: The JAM-derived mass volume density $\rho(R, z)$ (including both the stellar and DM components) is parameterized by Multi-Gaussian Expansion \citep[MGE;][]{Emsellem1994,Cappellari2002}, which can be integrated along the vertical ($z$) direction of the galaxy (with axisymmetric oblate assumption) to obtain the face-on surface mass density. The deprojected MGE mass density can be written as 
    \begin{equation}\label{eq:density_MGE}
        \rho(R, z) = \sum_{k=1}^{N} \frac{M_k}{(\sqrt{2\pi} \sigma_k)^{3} q_k} \exp\left[ -\frac{1}{2\sigma_k^{2}} \left( R^{2} + \frac{z^{2}}{q_k^{2}} \right) \right],
    \end{equation}
    where $N$ is the number of Gaussian components each with total mass $M_k$, dispersion $\sigma_k$ along the major axis of projected MGE, and intrinsic axial ratio $q_k$. Here, $q_k=\sqrt{q_{k}^{\prime2}-\cos^{2}{i}}/\sin{i}$, where $q_{k}^{\prime}$ is the observed axial ratio and $i$ is again the inclination angle. For the stellar component, $M_k=L_k\times (M_{\ast}/L)_{\rm dyn}$, where $L_k$ is the total luminosity of each Gaussian component and $(M_{\ast}/L)_{\rm dyn}$ is the best-fitting stellar mass-to-light ratio derived from the JAM model \citep[][keyword: \texttt{log\_ML\_stellar} in Table~B1]{Zhu2023}. For the DM component assuming spherical symmetry (e.g. the gNFW halo adopted in this work), the density expressed in the form of \autoref{eq:density_MGE} is obtained by fitting a one-dimensional MGE model to the analytic parameterization. Then the face-on total mass surface density is obtained by integrating the sum of two sets of MGEs (stellar density $\rho_{\ast}$ and DM density $\rho_{\rm DM}$) along the z-axis, written as
    \begin{equation}
        \Sigma_{\rm total}(R) = \int_{z_{\min}}^{z_{\max}} \left[\rho_{\ast}(R, z)+\rho_{\rm DM}(R, z)\right] \, {\rm d}z. 
    \end{equation}
    To avoid including the DM distribution located far away from the galaxy that is unlikely to affect the star formation activities, we adopt $z_{\min}=-3\,\max(\sigma_{k})\,q_{\rm e, intr}$ and $z_{\max}=3\,\max(\sigma_{k})\,q_{\rm e, intr}$. Here $\max(\sigma_{k})$ is the maximum dispersion of the luminous Gaussians along the major axis, $q_{\rm e, intr}$ is the intrinsic minor-to-major axial ratio defined as $q_{\rm e, intr}\equiv \sqrt{(1-\varepsilon_{\rm e})^2-\cos^2{i}}/\sin{i}$, where $\varepsilon_{\rm e}$ is the observed ellipticity of a half-light elliptical isophote derived from MGE (keyword: \texttt{Eps\_MGE}) and $i$ is the inclination angle.

    \item Circular velocity $V_{\rm circ}$: Following \citet{Zhu2025}, we assume that the galaxies are axisymmetric and use the \textsc{jampy.mge.vcirc}\footnote{Version 8.1.4 in \url{https://pypi.org/project/jampy/}} routine \citep[][Eq.~45]{Cappellari2020} to derive maps with constant $V_{\rm circ}$ at a given face-on projected galactocentric radius $R=\sqrt{x_{\rm bin}^2 + y_{\rm bin}^2/(1-\varepsilon_{\rm e})^{2}}$, where ($x_{\rm bin}$, $y_{\rm bin}$) are the coordinates of Voronoi bins and $\varepsilon_{\rm e}$ (keyword: \texttt{Eps\_MGE}) is the observed ellipticity of a half-light elliptical isophote derived from the MGE model.
    
    \item Oort parameter $A_{\rm Oort}$: Galactic shear is quantified using the first Oort parameter $A_{\rm Oort}$ \citep{Oort1927}, defined as
    \begin{equation}
    A(R)\equiv\left|\frac{1}{2}\left(\frac{V_{\rm circ}}{R}-\frac{{\rm d}V_{\rm circ}}{{\rm d}R}\right)\right|=\left|-0.5R\frac{{\rm d}\Omega}{{\rm d}R}\right|,
    \end{equation}
    where $V_{\rm circ}$ and $\Omega\equiv V_{\rm circ}/R$ are the circular and angular velocities at a given galactocentric radius $R$ in the equatorial plane. We compute $\frac{{\rm d}V_{\rm circ}}{{\rm d}R}$ by analytically differentiating the MGE expression for the circular velocity \citep[][Eq.~45]{Cappellari2020}. This avoids numerical differentiation of the circular velocity curve and yields a smooth and robust estimate of the velocity gradient. The resulting velocity gradient is also used to derive the epicyclic frequency, which is defined as
    \begin{equation}\label{eq:epicyclic_frequency}
        \kappa=\sqrt{R\frac{{\rm d}\Omega^2}{{\rm d}R}+4\Omega^2}=\sqrt{\frac{2V_{\rm circ}}{R}\left(\frac{{\rm d}V_{\rm circ}}{{\rm d}R}+\frac{V_{\rm circ}}{R}\right)}.
    \end{equation}
    Note that we adopt the absolute values of $A_{\rm Oort}$, which differs slightly from that given by \citet[Eq.~3.83]{Binney2008}, to account only for the strength of shear.

    \item Disk instability parameter $Q_{\rm Toomre}$: We use the modified version of the \citet{Toomre1964} $Q$ parameter to estimate the ability of a disk with both gas and stars to stabilize itself \citep{Wang1994}, defined as
    \begin{equation}\label{eq:ToomreQ}
        Q_{\rm Toomre}= \left(\frac{1}{Q_{\rm gas}}+\frac{1}{Q_{\ast}}\right)^{-1}.
    \end{equation}
    Here, $Q_{\rm gas}$ and $Q_{\ast}$ are the gravitational instability parameters for the gas and stellar components, respectively, defined as
    \begin{equation}
        Q_{\rm gas}=\frac{\kappa \sigma_{\rm R, gas}}{\pi G \Sigma_{\rm gas}}; Q_{\ast}=\frac{\kappa \sigma_{\rm R, \ast}}{\pi G \Sigma_{\ast}},
    \end{equation}
    where $\kappa$ is the epicyclic frequency derived from \autoref{eq:epicyclic_frequency}, $\Sigma_{\rm gas}$ is the total gas surface density including molecular and atomic gas, $\sigma_{\rm R, gas}$ is the radial gas velocity dispersion, and $\sigma_{\rm R, \ast}$ is the radial stellar velocity dispersion that can be computed using \textsc{jampy.axi.intr} for a given best-fit JAM model (assuming a spatially constant velocity anisotropy in our case). For the radial gas velocity dispersion of each bin, we adopt the line-of-sight velocity dispersion $\sigma_{\rm H_2}$ maps or a constant value of the characteristic sound speed of $10^4$ K gas \citep[$c_{\rm s}=13\,{\rm km\,s^{-1}}$;][]{Stark2018}, and our results are consistent under these two assumptions. To account for the contribution of the DM potential on stabilizing gas instability, we also replace $\Sigma_{\ast}$ (from SPS model) with $\Sigma_{\rm total}-\Sigma_{\rm gas}$, which represents the surface density of collisionless components (including the stellar and DM), and the conclusions remain unchanged.

    \item Mid-plane dynamical equilibrium pressure $P_{\rm DE}$: $P_{\rm DE}$ (as defined in \autoref{eq:PDE}) reflects the balance between the weight of the interstellar medium (ISM) and the supporting forces such as energy/momentum input from the massive stars \citep[e.g.,][]{Ostriker2010,Ostriker2022}. Following \citet[][Eq.~12]{Sun2020b} and \citet[][Eq.~2]{Ellison2024}, $P_{\rm DE}$ is expressed as
    \begin{equation}\label{eq:PDE}
        P_{\rm DE}=\frac{\pi G}{2}\Sigma_{\rm gas}^2+\Sigma_{\rm gas}\sqrt{2G\rho_{\ast}}\sigma_{\rm z, gas},
    \end{equation}
    where $\rho_{\ast}$ is the mid-plane stellar mass volume density, and $\sigma_{\rm z, gas}$ is the gas velocity dispersion perpendicular to the disk. For each Voronoi bin with a face-on projected galactocentric radius $R$ (see the definition above), we analytically compute the luminosity density $\rho_{\rm lum} (R, z)$ (expressed in a form similar to \autoref{eq:density_MGE}, but using $L_{k}$ instead of $M_{k}$) in the mid-plane ($z=0$), and then multiply it by the dust-corrected stellar $M_{\ast}/L$ maps derived from SPS models \citep[][Table~B1 with keyword: \texttt{ML\_int\_Map}]{Lu2023} to obtain $\rho_{\ast}$. Using either a constant value of $\sigma_{\rm z, gas}=11\,{\rm km\,s^{-1}}$ \citep{Leroy2008} or the line-of-sight velocity dispersion $\sigma_{\rm z, gas}=\sigma_{\rm H_2}$ does not alter our results and conclusions.

\end{itemize}

\section{Results}\label{sec:results}
\subsection{2D maps of spatially resolved properties}\label{subsec:2Dmaps}
\begin{figure*}
    \centering
    \includegraphics[width=\textwidth]{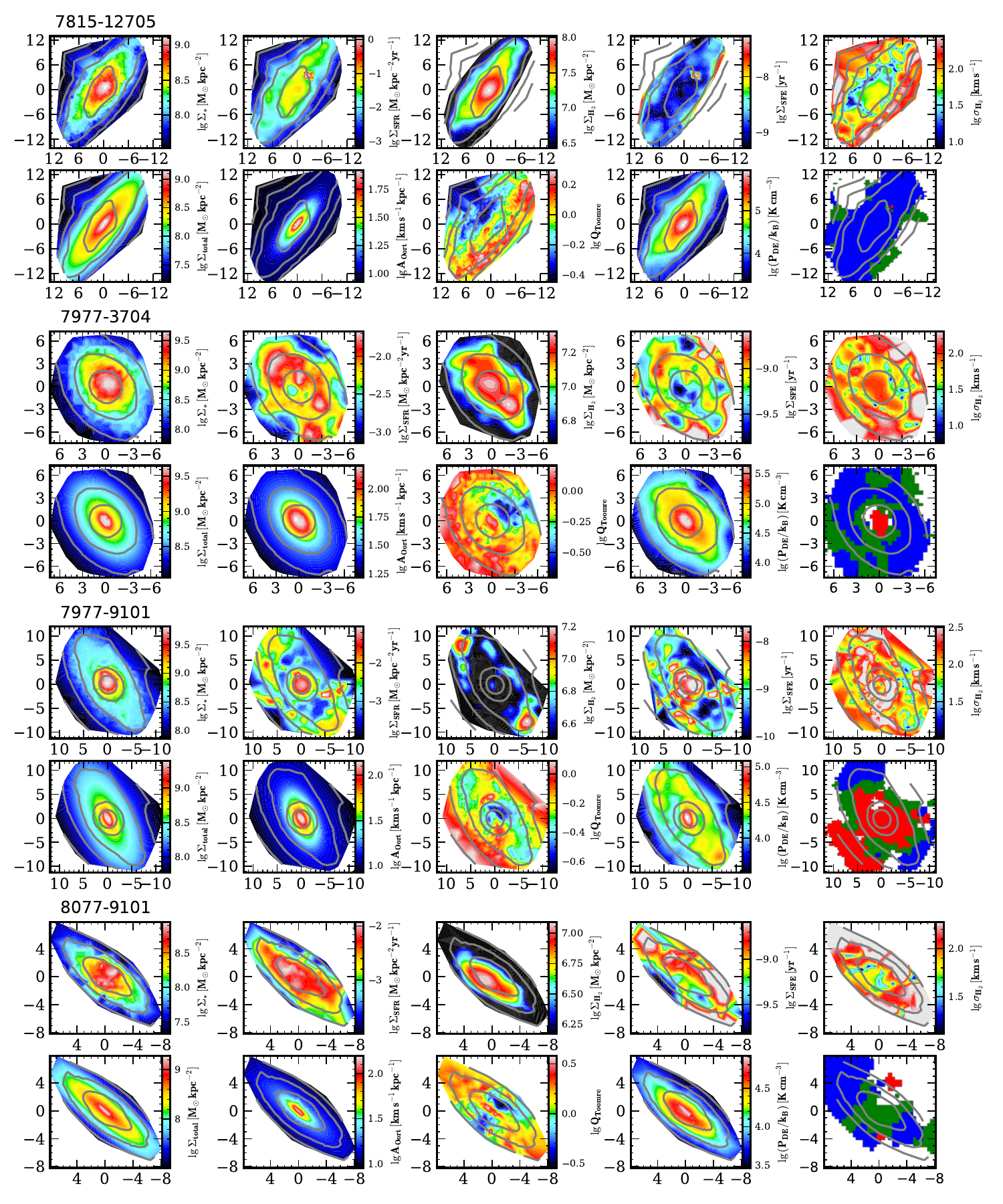}
    \caption{Resolved maps of key physical properties for ALMaQUEST galaxies (10 maps for each galaxy). \textit{Top row}: the panels from left to right show the stellar mass surface density $\Sigma_{\ast}$, SFR surface density $\Sigma_{\rm SFR}$, molecular gas surface density $\Sigma_{\rm H_2}$, surface density of SFE $\Sigma_{\rm SFE}$, gas velocity dispersion $\sigma_{\rm H_2}$. \textit{Bottom row}: the panels from left to right show the total mass surface density $\Sigma_{\rm total}$, Oort shear parameter $A_{\rm Oort}$, Toomre stability parameter $Q_{\rm Toomre}$, mid-plane dynamical equilibrium pressure $P_{\rm DE}$, and BPT classifications (red: AGN, green: composite, blue: SF). The gray contours are the observed surface brightness contours in SDSS $g$-band (in steps of 1 mag). The complete figure set (8 images) is available in the online article.}
    \label{fig:maps_example}
\end{figure*}

In \autoref{fig:maps_example}, we present 2D maps of spatially resolved properties for example galaxies. For each galaxy, we show several properties, including stellar mass surface density $\Sigma_{\ast}$, SFR surface density $\Sigma_{\rm SFR}$, molecular hydrogen surface density $\Sigma_{\rm H_2}$, star formation efficiency $\Sigma_{\rm SFE}$, velocity dispersion of molecular gas $\sigma_{\rm H_2}$, total mass surface density $\Sigma_{\rm total}$, Oort parameter $A_{\rm Oort}$, Toomre Q parameter $Q_{\rm Toomre}$, mid-plane dynamical equilibrium pressure $P_{\rm DE}/k_{\rm B}$, classification of regions (star-forming, AGN, and composite) based on the BPT diagram. Note that although we show all Voronoi bins in \autoref{fig:maps_example}, only star-forming regions are included in the following analysis. 

As shown in \autoref{fig:maps_example}, the $\Sigma_{\ast}$, $\Sigma_{\rm SFR}$, and $\Sigma_{\rm H_2}$ values correlate well with each other, following the resolved Kennicutt-Schmidt (rKS; $\Sigma_{\rm SFR}$ versus $\Sigma_{\rm H_2}$) relation, resolved star formation main sequence (rSFMS; $\Sigma_{\rm SFR}$ versus $\Sigma_{\ast}$), and resolved molecular gas main sequence (rMGMS; $\Sigma_{\rm H_2}$ versus $\Sigma_{\ast}$) that have been presented in previous papers of the ALMaQUEST series \citep{Lin2019,Ellison2021,Lin2022}. Most galaxies exhibit lower SFE values in their central regions, where stellar mass surface densities are typically higher. This trend in the ALMaQUEST sample has been reported by \citet{Lin2019}, who found a negative effect of $\Sigma_{\rm SFR} \propto (\Sigma_{\ast}^{-0.3}\Sigma_{\rm H_2})^{1.38}$. However, the dependence on $\Sigma_{\ast}$ is weak as the scatter of this relation is not significantly different from the original rKS relation. This result appears to be in tension with the extended rKS law $\Sigma_{\rm SFR} \propto \Sigma_{\ast}^{0.5}\Sigma_{\rm gas}$ \citep[proposed by][]{Shi2011,Shi2018}, in which the mid-plane gas pressure governed by stellar gravity plays a positive role in converting cold gas into stars. As discussed in \citet{Lin2019} and \citet[][Section~4.1]{Jing2024}, the discrepancy is probably due to differences in the sample properties, such as the range of stellar mass surface density or gas fraction \citep[also see the dependence of dynamical suppression on gas fraction in][]{Gensior2021}. Moreover, at fixed $\Sigma_{\rm gas}$, \citet{Sanchez2021} reported a positive role of $\Sigma_{\ast}$ in regulating SFR in the EDGE-CALIFA survey, while \citet[][Figure~6]{Pessa2022} found in the PHANGS galaxies that the role of $\Sigma_{\ast}$ can vary from negative to positive for different spatial resolutions (from 150 pc to kpc) and different galactic environments (e.g., bar, ring, disk, and spiral arm).

The role of stellar mass surface density is usually thought to be related to the gravitational potential, although this connection is not yet fully clear. However, the stellar component may not perfectly trace the total gravitational potential, as the contribution from the DM halo can not be ignored in some cases within our sample (see \autoref{fig:galprop}). As shown in the bottom two panels for each galaxy in \autoref{fig:maps_example}, quantities related to the gravitational potential, namely the total mass surface density $\Sigma_{\rm total}$ and the galactic shear $A_{\rm Oort}$, increase towards galactic center. Although these trends are qualitatively similar to that of $\Sigma_{\ast}$ based on visual inspection of the 2D maps, stellar mass surface density alone may not fully capture the underlying gravitational effects. It is therefore necessary to assess the relative importance of these different quantities in regulating star formation.

The \citet{Toomre1964} stability parameter, $Q_{\rm Toomre}$ (defined in \autoref{eq:ToomreQ}), quantifies the susceptibility of a galactic disk to gravitational instability, with lower values corresponding to a more unstable gas disk. Such instabilities are expected to promote the formation of dense molecular structures that can subsequently collapse to form stars. Consistent with this expectation, regions with lower $Q_{\rm Toomre}$ generally exhibit higher $\Sigma_{\rm SFR}$ in most galaxies. In contrast, the relationship between $Q_{\rm Toomre}$ and $\Sigma_{\rm SFE}$ is weaker.

The mid-plane dynamical equilibrium pressure, $P_{\rm DE}$, has also been found to correlate positively with $\Sigma_{\rm SFR}$ across several molecular gas surveys, including \citet[][PHANGS]{Sun2020b}, \citet[][EDGE-CALIFA]{Barrera-Ballesteros2021}, and \citet[][ALMaQUEST]{Ellison2024}, as well as in the star-forming disk of M33 \citep{Corbelli2025}. As illustrated in \autoref{fig:maps_example}, regions with higher $P_{\rm DE}$ (typically located near galaxy centers) generally correspond to enhanced $\Sigma_{\rm SFR}$, in agreement with previous studies. For $\Sigma_{\rm SFE}$, however, we observe an anti-correlation with $P_{\rm DE}$. This trend may arise because $P_{\rm DE}$ primarily traces the fraction of the ISM in the molecular phase rather than the efficiency with which molecular gas is converted into stars \citep[e.g.,][]{Leroy2008}. Furthermore, the apparent anti-correlation between $\Sigma_{\rm SFE}$ and $P_{\rm DE}$ may be influenced by covariances with other physical quantities. For example, both $P_{\rm DE}$ and $Q_{\rm Toomre}$ depend on gas and stellar densities, introducing strong inter-dependencies among these parameters. Such couplings often produce similar spatial patterns in the 2D distributions of different physical quantities, making it difficult to determine the primary drivers of SFE variations through qualitative comparisons of 2D maps alone.

\subsection{Random forest regression results}\label{subsec:RF}
To quantify the relative importance of different physical parameters in regulating SFE, we employ a random forest regression model. A random forest is an ensemble learning method that combines a large number of decision trees, where each tree recursively partitions the parameter space into subsets with similar target values. During training, individual trees are constructed using random subsets of both the data and input features, which reduces overfitting and improves predictive performance. The final prediction is obtained by averaging the predictions from all trees, allowing the model to capture complex and non-linear relationships between SFE and its potential drivers. Importantly, random forest provides estimates of feature importance, enabling us to assess the relative contribution of different physical parameters to the observed variations in SFE. We refer readers to \citet[][Appendix B]{Bluck2022} for more details about the random forest methodology. 

For the practical implementation, we utilize the \texttt{RandomForestRegressor} function from the powerful Python package \textsc{scikit-learn} \citep{Pedregosa2011}. We investigate which local quantities (measured on Voronoi bins) are most predictive of $\Sigma_{\rm SFE}$ among the following: stellar mass surface density $\Sigma_{\ast}$, total mass surface density $\Sigma_{\rm total}$, galactic shear $A_{\rm Oort}$, stability parameter $Q_{\rm Toomre}$, mid-plane dynamical equilibrium pressure $P_{\rm DE}$, and gas velocity dispersion $\sigma_{\rm H_2}$. For reference, we also include a random sequence in the random forest analysis. Owing to the limited sample size of this work, we do not include global galaxy properties in the analysis, which have been shown to have weak effects on local star formation \citep[e.g.,][]{Bluck2022,Jing2024}. 

The data are split into a training set (70$\%$ of the data) and a test set (30$\%$ of the data), both consisting of the predictor quantities paired with the target quantity ($\Sigma_{\rm SFE}$). The random forest regression is fit on the training data, and then applied to the test data to evaluate its predictive performance using the mean squared error (MSE),
\begin{equation}
    {\rm MSE} = \frac{\sum_{i}^{N}(y_{{\rm true},i}-y_{{\rm pred},i})^2}{N},
\end{equation}
and the coefficient of determination ($R^2$),
\begin{equation}
    R^2 = 1-\frac{\sum_{i}^{N}(y_{{\rm true},i}-y_{{\rm pred},i})^2}{\sum_{i}^{N}(y_{{\rm true},i}-\bar{y}_{\rm true})^2}.
\end{equation}
Here, $N$ is the number of data points, $y_{{\rm pred},i}$ and $y_{{\rm true},i}$ are the predicted and true (measured) values of the ith data points, respectively. $\bar{y}_{\rm true}$ is the mean value of the true data. We use the \texttt{GridSearchCV} module in \textsc{scikit-learn} to tune the hyperparameters and adopt a threshold on test-to-training MSE ratio ($\rm MSE_{test}/MSE_{train}\leqslant1.5$) to mitigate overfitting. Adopting different thresholds (1.2 or 2.0) does not qualitatively affect the results. The fine-tuned hyperparameters are: \texttt{n\_estimators = 200}, \texttt{max\_depth = 25}, \texttt{min\_samples\_split = 30}, and \texttt{min\_samples\_leaf = 5}.

\begin{figure*}
    \centering
    \includegraphics[width=\textwidth]{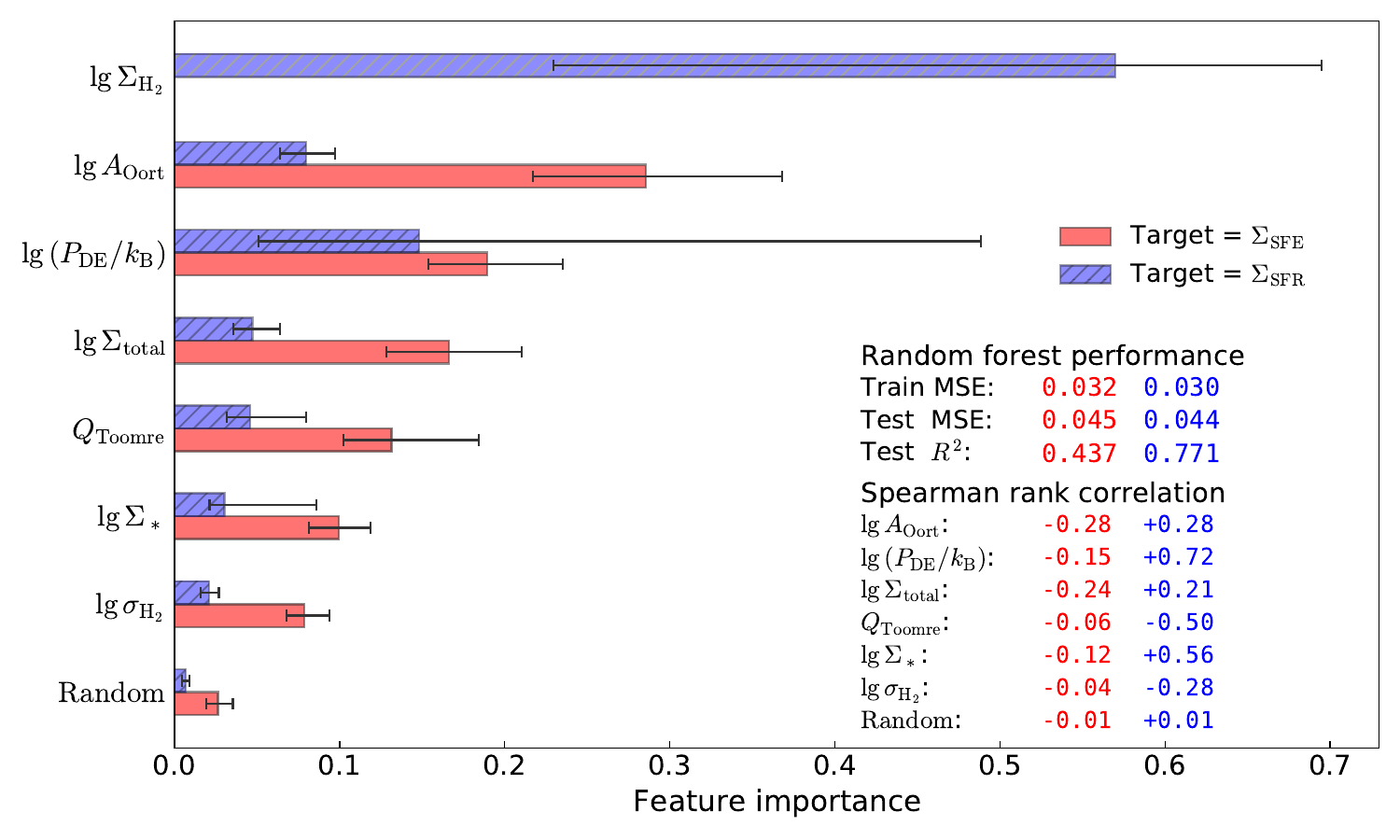}
    \caption{Feature importance derived from random forest regression models predicting $\Sigma_{\rm SFE}$ (red) and $\Sigma_{\rm SFR}$ (blue hatched). For each feature, the bars show the median importance computed from individual decision trees in the forest, while the error bars indicate the 16th–84th percentile range. From top to bottom, the seven features are ranked by decreasing median importance (for the $\Sigma_{\rm SFE}$ prediction), with an additional feature ($\Sigma_{\rm H_2}$) included in the analysis targeting $\Sigma_{\rm SFR}$. The inset text summarizes the random forest performance for each model, including the training and test mean squared errors (MSE) and the test $R^2$, evaluated using a 70/30 train–test split. The lower text block lists the Spearman rank correlation coefficients between $\Sigma_{\rm SFE}$ and each feature for the two models.}
    \label{fig:RF}
\end{figure*}

The results of the random forest regression are summarized in \autoref{fig:RF}. Based on the relative feature importance derived from the model targeted on SFE (red bar charts), the seven quantities are ranked from top to bottom, as shown in the accompanying bar charts. The model exhibits moderate predictive performance, with a coefficient of determination of $R^2=0.437$, considering only six physically meaningful input quantities (except for the random variable). Although the random forest model explains only a fraction of the total variance in the target variable, its primary purpose here is to compare the relative predictive power of different physical parameters rather than to construct an optimal predictive model. We further assess the robustness of the inferred feature rankings using permutation importance, which measures the degradation in predictive performance ($R^2$) when a given feature is randomly shuffled. The rankings obtained from permutation importance are identical to those derived from the feature importance, suggesting that the relative importance of the considered physical parameters is reasonably stable and not strongly dependent on the specific importance estimator adopted. The ratio of test-to-training MSE remains at an acceptable level, indicating no substantial overfitting. Notably, all computed Spearman rank correlation coefficients $\rho_{\rm Spearman}$ between $\Sigma_{\rm SFE}$ and the other quantities are negative, suggesting a systematic anti-correlation across the sample.

The random forest model identifies galactic shear, quantified by the Oort parameter $A_{\rm Oort}$, as the most important quantity in predicting $\Sigma_{\rm SFE}$, with an anti-correlation of $\rho_{\rm Spearman}=-0.28$. It should be noted that the feature importance derived by the random forest method only denotes correlation but not necessarily causation. However, given that $\Sigma_{\rm SFE}$ traces present-day star-formation activity, which does not directly modify the current gravitational potential (more closely related to the integrated mass-growth history), the observed anti-correlation is most plausibly attributed to the suppressive role of galactic shear in regulating SFE.

Notably, the total mass surface density $\Sigma_{\rm total}$, another indicator of the gravitational potential, plays a relatively minor role in regulating $\Sigma_{\rm SFE}$ compared to $A_{\rm Oort}$, although it remains slightly more important than $\Sigma_{\ast}$. One plausible explanation is that, for a given gravitational potential $\Phi(R, z)$, the local $\Sigma_{\rm total}$ (integrated along the $z$-axis) primarily captures information in the vertical direction \citep[e.g., the mid-plane pressure associated with vertical gravity, as discussed in][]{Ostriker2010,Kim2011}. Consequently, $\Sigma_{\rm total}$ may not fully encapsulate dynamical effects in the radial direction, such as differential rotation. In contrast, the galactic shear $A_{\rm Oort}$ is related both to the local mass surface density at radius $R$ and to the average mass surface density enclosed within $R$ \citep[see Eq.~5 in][]{Jing2024}, thereby providing a more complete description of the radially varying gravitational potential that influences SFE.

The mid-plane dynamical equilibrium pressure, $P_{\rm DE}$, is less important than galactic shear in predicting $\Sigma_{\rm SFE}$, but plays a more prominent role in predicting $\Sigma_{\rm SFR}$. This difference may partly reflect the fact that $\Sigma_{\rm SFR}$ is an extensive quantity, whereas $\Sigma_{\rm SFE}$ is a normalized quantity. Consequently, a strong correlation with $\Sigma_{\rm SFR}$ does not necessarily imply an equally strong correlation with $\Sigma_{\rm SFE}$. Nevertheless, the relative importance rankings still suggest that the physical processes governing SFR and SFE may not be identical. While $P_{\rm DE}$ is a stronger predictor of $\Sigma_{\rm SFR}$, consistent with its role in regulating the molecular gas content of the ISM, galactic shear emerges as the most important predictor of $\Sigma_{\rm SFE}$. This result suggests that local galactic dynamics, particularly shear, may play a more significant role than pressure in determining the efficiency with which molecular gas is converted into stars. Interestingly, \citet{Lin2024} found that the CO-based $\Sigma_{\rm SFE}$ is tightly correlated with the dense-gas SFE traced by HCN and HCO$^+$. Future studies of the dependence of dense-gas SFE on galactic shear could therefore provide a more direct probe of the role of shear in star formation.

As shown in \autoref{fig:RF}, however, the importance of $P_{\rm DE}$ for $\Sigma_{\rm SFR}$ remains substantially lower than that of the molecular gas surface density, $\Sigma_{\rm H_2}$. This differs from the random forest analysis of \citet[][Figure~10]{Ellison2024}, who found $P_{\rm DE}$ to be significantly more important than $\Sigma_{\rm H_2}$ in predicting $\Sigma_{\rm SFR}$. A likely explanation is the difference in sample selection (see \autoref{subsec:manga} and \autoref{fig:galprop}) between the two studies. Our sample excludes merging systems in order to ensure reliable dynamical modeling, whereas \citet{Ellison2024} analyzed a larger sample that includes a substantial fraction of merging and starburst galaxies. In addition, the relatively large uncertainties associated with the feature importances of both $\Sigma_{\rm H_2}$ and $P_{\rm DE}$ may reflect the strong covariance between these quantities, given that molecular gas surface density enters directly into the definition of $P_{\rm DE}$ (see \autoref{eq:PDE}). Indeed, we find that $P_{\rm DE}$ becomes the dominant predictor of $\Sigma_{\rm SFR}$, with a median feature importance of $\sim0.7$, when $\Sigma_{\rm H_2}$ is excluded from the random forest analysis. To further test the effect of sample selection, we repeat the analysis using the same galaxy sample and input variables as \citet{Ellison2024}. In this case, $P_{\rm DE}$ again emerges as the dominant predictor, with a feature importance of $\sim0.8$, in excellent agreement with \citet[][Figure~10]{Ellison2024}. This result strongly suggests that differences in sample selection, rather than the random forest methodology itself, are the primary source of the discrepancy between our results and those of \citet{Ellison2024}.

Before summarizing the random forest results, we examine the robustness of the feature rankings to several measurement and sample-selection choices. These tests are summarized in \autoref{tab:rf_importance_models}. First, the analysis using azimuthally averaged quantities is presented in \autoref{appendix:rf_azimuthal}. Briefly, $A_{\rm Oort}$ remains an important predictor after azimuthal averaging, although its relative importance is reduced. Second, we test the sensitivity of $A_{\rm Oort}$ to the adopted rotation curve. Varying the MaNGA PSF FWHM by $\pm10\%$ in the JAM modelling produces only modest changes in the inferred circular velocity curves and the resulting feature importances (see \autoref{appendix:cvc_comparison} and \autoref{tab:rf_importance_models}). The CO-based rotation curves show some differences in the inner regions, but $A_{\rm Oort}$ remains the highest-ranked predictor when calculated instead from the CO-based rotation curves (\autoref{tab:rf_importance_models}). Third, removing the BPT selection, including $R/R_{\rm e}$ as an additional predictor, and separately analysing regions inside and outside $R_{\rm e}$ do not remove the preference for $A_{\rm Oort}$. In particular, its importance remains higher than that of $R/R_{\rm e}$ when both are included, indicating that its predictive importance cannot be explained solely by their shared radial dependence.

The feature rankings may also depend on the adopted tracers of $\Sigma_{\rm H_2}$, SFR, and $\Sigma_{\ast}$. For $\Sigma_{\rm H_2}$, adopting a metallicity-dependent $\alpha_{\rm CO}$ \citep{Sun2020a} produces no qualitative change in the feature ranking (\autoref{tab:rf_importance_models}). Previous studies have used hybrid UV--IR or H$\alpha$--IR SFR indicators and photometric near-infrared estimates of $\Sigma_{\ast}$ \citep[e.g.,][]{Leroy2008,Leroy2013,Sun2023}. Using Balmer-decrement-corrected H$\alpha$ emission as a reference, \citet{Belfiore2023} found that traditional hybrid recipes tend to overestimate the SFR in regions with low $\Sigma_{\rm SFR}$ and low specific SFR, whereas their recalibrated recipes agree with the reference SFR to within approximately 0.2 dex on resolved scales, with negligible systematic bias. We therefore do not expect the use of alternative, well-calibrated SFR tracers to qualitatively alter our conclusions. Meanwhile, photometric estimates of $\Sigma_{\ast}$ based on Wide-field Infrared Survey Explorer (WISE) W1 or Spitzer $3.6\,\mu{\rm m}$ imaging \citep[e.g.,][]{Leroy2013,Sun2023} are not uniformly available for the selected sample at the matched spatial resolution required here. Testing these alternative measurements would therefore require degrading the full analysis to a substantially coarser physical scale, thereby reducing both the available spatial information and the number of resolved regions. The quantitative impact of adopting photometric $\Sigma_{\ast}$ estimates on the feature importances therefore remains a limitation of the present work.

Overall, these tests show that the principal feature rankings are broadly robust to the modelling and sample-selection choices considered here. A notable exception occurs when each quantity is normalized by its within-galaxy median: in this case, $P_{\rm DE}$ becomes the leading predictor, while the relative importance of $A_{\rm Oort}$ decreases. This suggests that galaxy-to-galaxy variations contribute to the prominence of $A_{\rm Oort}$ in the fiducial regression. At the same time, the fact that $A_{\rm Oort}$ retains non-negligible predictive importance after within-galaxy normalization indicates that its contribution is not driven exclusively by global differences among galaxies. 

\begin{deluxetable*}{@{}lccccccc@{}}
\tabletypesize{\scriptsize}
\tablewidth{\textwidth}
\tablecaption{Median random forest feature importances for the fiducial model and robustness tests.\label{tab:rf_importance_models}}
\tablehead{
\colhead{Model} & \colhead{$\Sigma_{\star}$} & \colhead{$\Sigma_{\rm total}$} & \colhead{$A_{\rm Oort}$} & \colhead{$P_{\rm DE}/k_{\rm B}$} & \colhead{$Q_{\rm Toomre}$} & \colhead{$\sigma_{\rm H_2}$} & \colhead{Random}
}
\startdata
(1) Fiducial & $0.100_{-0.018}^{+0.019}$ & $0.166_{-0.038}^{+0.044}$ & $\mathbf{0.285_{-0.068}^{+0.082}}$ & $0.190_{-0.036}^{+0.046}$ & $0.132_{-0.029}^{+0.052}$ & $0.079_{-0.011}^{+0.015}$ & $0.026_{-0.007}^{+0.009}$ \\
(2) Metallicity-dependent $\alpha_{\rm CO}$ & $0.123_{-0.020}^{+0.019}$ & $0.183_{-0.022}^{+0.023}$ & $\mathbf{0.277_{-0.026}^{+0.027}}$ & $0.175_{-0.023}^{+0.021}$ & $0.130_{-0.016}^{+0.015}$ & $0.085_{-0.016}^{+0.015}$ & $0.027_{-0.009}^{+0.009}$ \\
(3) DynPop: $0.9\,{\rm FWHM}$ & $0.101_{-0.020}^{+0.021}$ & $0.162_{-0.035}^{+0.046}$ & $\mathbf{0.287_{-0.057}^{+0.076}}$ & $0.182_{-0.029}^{+0.041}$ & $0.132_{-0.025}^{+0.056}$ & $0.081_{-0.013}^{+0.016}$ & $0.031_{-0.008}^{+0.008}$ \\
(4) DynPop: $1.1\,{\rm FWHM}$ & $0.103_{-0.020}^{+0.023}$ & $0.160_{-0.035}^{+0.043}$ & $\mathbf{0.285_{-0.056}^{+0.075}}$ & $0.188_{-0.029}^{+0.038}$ & $0.131_{-0.025}^{+0.048}$ & $0.081_{-0.015}^{+0.019}$ & $0.030_{-0.008}^{+0.008}$ \\
(5) CO RC: URC & $0.126_{-0.024}^{+0.028}$ & $0.205_{-0.048}^{+0.038}$ & $\mathbf{0.251_{-0.037}^{+0.036}}$ & $0.182_{-0.027}^{+0.027}$ & $0.108_{-0.023}^{+0.025}$ & $0.095_{-0.019}^{+0.019}$ & $0.031_{-0.011}^{+0.014}$ \\
(6) CO RC: cubic B-spline & $0.126_{-0.022}^{+0.021}$ & $0.155_{-0.027}^{+0.030}$ & $\mathbf{0.294_{-0.031}^{+0.033}}$ & $0.177_{-0.025}^{+0.023}$ & $0.099_{-0.019}^{+0.023}$ & $0.108_{-0.019}^{+0.021}$ & $0.035_{-0.012}^{+0.014}$ \\
(7) CO-common: DynPop & $0.094_{-0.021}^{+0.031}$ & $0.144_{-0.026}^{+0.027}$ & $\mathbf{0.311_{-0.028}^{+0.028}}$ & $0.178_{-0.020}^{+0.019}$ & $0.128_{-0.023}^{+0.020}$ & $0.103_{-0.019}^{+0.020}$ & $0.036_{-0.012}^{+0.016}$ \\
(8) No BPT selection & $0.105_{-0.013}^{+0.012}$ & $0.175_{-0.014}^{+0.013}$ & $\mathbf{0.414_{-0.019}^{+0.021}}$ & $0.120_{-0.011}^{+0.012}$ & $0.091_{-0.012}^{+0.014}$ & $0.081_{-0.009}^{+0.009}$ & $0.014_{-0.004}^{+0.004}$ \\
(9) No BPT selection + $R/R_{\rm e} (0.147_{-0.029}^{+0.024})$ & $0.115_{-0.037}^{+0.059}$ & $0.116_{-0.025}^{+0.041}$ & $\mathbf{0.309_{-0.050}^{+0.082}}$ & $0.098_{-0.013}^{+0.016}$ & $0.093_{-0.023}^{+0.039}$ & $0.072_{-0.013}^{+0.015}$ & $0.014_{-0.003}^{+0.004}$  \\
(10) No BPT selection: $R<R_{\rm e}$ & $0.070_{-0.014}^{+0.017}$ & $0.098_{-0.013}^{+0.014}$ & $\mathbf{0.522_{-0.025}^{+0.022}}$ & $0.108_{-0.012}^{+0.014}$ & $0.094_{-0.015}^{+0.017}$ & $0.091_{-0.012}^{+0.013}$ & $0.013_{-0.004}^{+0.005}$ \\
(11) No BPT selection: $R\geq R_{\rm e}$ & $0.108_{-0.033}^{+0.031}$ & $0.248_{-0.025}^{+0.023}$ & $\mathbf{0.340_{-0.036}^{+0.034}}$ & $0.132_{-0.017}^{+0.020}$ & $0.080_{-0.027}^{+0.043}$ & $0.057_{-0.015}^{+0.051}$ & $0.016_{-0.006}^{+0.007}$ \\
(12) Galaxy-normalized & $0.112_{-0.027}^{+0.039}$ & $0.147_{-0.027}^{+0.023}$ & $0.140_{-0.032}^{+0.110}$ & $\mathbf{0.291_{-0.065}^{+0.119}}$ & $0.090_{-0.021}^{+0.019}$ & $0.107_{-0.027}^{+0.089}$ & $0.041_{-0.012}^{+0.015}$ \\
\enddata
\tablecomments{
The fiducial model (1) is trained on the BPT-selected star-forming Voronoi bins and uses $\Sigma_{\star}$, $\Sigma_{\rm total}$, $A_{\rm Oort}$, $P_{\rm DE}/k_{\rm B}$, $Q_{\rm Toomre}$, $\sigma_{\rm H_2}$, and a random control variable as predictors of the SFE.
Relative to the fiducial model:
(2) adopts a metallicity-dependent CO-to-H$_2$ conversion factor;
(3) and (4) calculate $A_{\rm Oort}$ from the DynPop circular velocity curves obtained by adopting 0.9 and 1.1 times the measured MaNGA PSF FWHM, respectively, in the JAM modelling;
(5)--(7) restrict the analysis to the same 15 galaxies with available CO rotation curves \citep{Su2022} and calculate $A_{\rm Oort}$ using, respectively, a Universal Rotation Curve \citep[URC;][]{Persic1996} fit to the CO rotation curve, a regularized cubic B-spline fit to the CO rotation curve, and the fiducial DynPop circular velocity curve;
(8) includes all valid Voronoi bins without applying the BPT star-forming selection;
(9) uses the same sample as model (8) but additionally includes
$R/R_{\rm e}$ as a predictor, whose importance is $0.147_{-0.029}^{+0.024}$;
(10) and (11) use the sample without the BPT selection but restrict it to regions at $R<R_{\rm e}$ and $R\geq R_{\rm e}$, respectively; and
(12) repeats the fiducial regression after normalizing each physical quantity by its within-galaxy median, thereby reducing the contribution from global galaxy-to-galaxy variations.
Each entry gives the median tree-level mean-decrease-in-impurity feature importance. The lower and upper uncertainties correspond to $I_{\rm med}-P_{16}$ and $P_{84}-I_{\rm med}$, respectively, and therefore describe tree-to-tree scatter rather than galaxy-sampling uncertainty. Boldface identifies the feature with the largest median importance in each model.
}
\end{deluxetable*}

In summary, our fiducial analysis identifies the local galactic shear, $A_{\rm Oort}$, as the most important dynamical parameter associated with variations in $\Sigma_{\rm SFE}$. Nevertheless, the differences in feature importance among the highest-ranked parameters remain modest, and the random forest model achieves only moderate predictive performance ($R^2=0.437$). When each quantity is normalized by its median value within each galaxy, $P_{\rm DE}$ becomes the highest-ranked parameter, while $A_{\rm Oort}$ retains a non-negligible importance. This result indicates that the feature ranking in the fiducial analysis is partly influenced by galaxy-to-galaxy variations and further motivates examining the effect of shear on SFE on a galaxy-by-galaxy basis. Indeed, previous studies have reported positive, negative, or negligible correlations between galactic shear and star formation activity (see \autoref{sec:intro}), which may weaken the average trend across the full galaxy sample.

\subsection{Galaxy-to-galaxy variations in the role of shear in regulating SFE}\label{subsec:gal2gal_variation}
\begin{figure*}
    \centering
    \includegraphics[width=\textwidth]{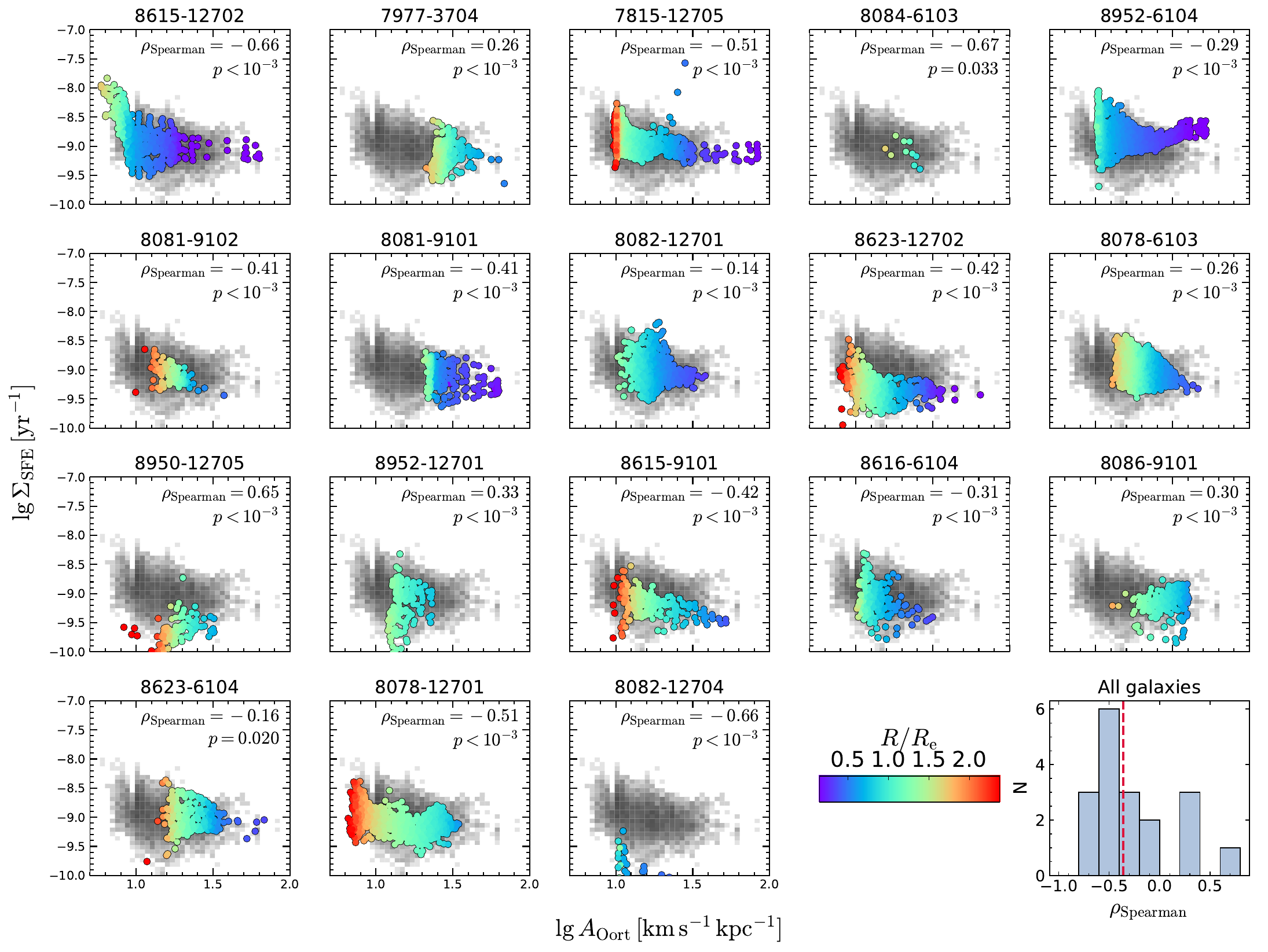}
    \caption{Resolved $\lg\,\Sigma_{\rm SFE}$–$\lg\,A_{\rm Oort}$ relations for individual galaxies with significant correlations ($p<0.05$), ordered by increased stellar mass. Each panel shows one galaxy, with points representing star-forming Voronoi bins color-coded by the normalized galactocentric radius ($R/R_{\rm e}$ with $R_{\rm e}$ is the circularized half-light radius). The gray background map represents the number density distribution of the 7925 star-forming Voronoi bins from all galaxies. The Spearman rank correlation coefficient and $p$-value are given in each panel. The histogram in the bottom-right panel shows the distribution of Spearman rank correlation coefficients for the 18 selected galaxies, with the median value indicated by the red vertical dashed line. A wide diversity in both the strength and the sign of the correlation is observed among galaxies.}
    \label{fig:shear_SFE_individual}
\end{figure*}

To examine whether the anti-correlation between SFE and galactic shear represents a unified relation across all galaxies, we investigate the $\Sigma_{\rm SFE}$–$A_{\rm Oort}$ relations for individual galaxies in \autoref{fig:shear_SFE_individual}. Each panel displays the corresponding Spearman’s rank correlation coefficient $\rho_{\rm Spearman}$ and its $p$-value. Of the 29 galaxies in the sample, two (plateifu: 8155-6101 and 8728-3701) are excluded because they contain no Voronoi bins meeting the selection criteria described in \autoref{subsec:spatial_prop}. Among the remaining 27 galaxies, 18 galaxies with statistically significant correlations (i.e., $p<0.05$) are presented in \autoref{fig:shear_SFE_individual}. Most of these galaxies (14 out of 18) exhibit anti-correlations ($\rho_{\rm Spearman}<0$) between SFE and galactic shear, with some displaying even stronger anti-correlations than the overall trend ($\rho_{\rm Spearman}=-0.28$ shown in \autoref{fig:RF}).

The strength of the SFE–shear anti-correlation varies substantially from galaxy to galaxy, indicating that galactic shear does not regulate SFE in a strictly universal manner. Moreover, four galaxies show a positive correlation between galactic shear and SFE ($\rho_{\rm Spearman}>0$), further confirming galaxy-to-galaxy variations in the effects of galactic shear on star formation activity. One possible origin of this variation is the presence of competing shear-related mechanisms.

According to the cloud–cloud collision model \citep[][Eqs.~12–14]{Tan2000}, the SFR surface density is hypothesized to be inversely proportional to the collision timescale $t_{\rm coll}$, such that $\Sigma_{\rm SFR}\propto \Sigma_{\rm gas}/t_{\rm coll}$. The collision timescale scales as $t_{\rm coll} \propto 1/(n,v,\sigma_{\rm coll}) \propto 1/\Sigma_{\rm gas}$, where $n$ is the number density of clouds, $v$ is their relative velocity, and $\sigma_{\rm coll}$ is the average cloud cross section. As gas density increases, the collision rate between GMCs rises, enhancing the efficiency of cloud–cloud collisions and strengthening the dependence of SFE on shear. 

This effect competes with the stabilizing influence of galactic shear and the background gravitational potential, which can inhibit the gravitational collapse of molecular gas and thereby reduce SFE. The framework proposed by \citet{Meidt2025} provides a related physical interpretation of this stabilizing effect: lower-density gas in cloud envelopes can remain kinematically coupled to motions generated by the host-galaxy potential, delaying its transition to self-gravitating collapse and limiting the
replenishment of the dense, star-forming component. In this picture, $A_{\rm Oort}$ traces the radial differential rotation component of the broader dynamical influence of the galactic potential, although other orbital, tidal, vertical, and non-circular motions may also contribute to the coupling.

The dependence of SFE on shear may therefore vary with gas density because of the competition between these two effects. In low-density regions, the stabilizing influence of shear and the background potential dominates, resulting in a stronger negative correlation between SFE and $A_{\rm Oort}$. In high-density regions, the increasing importance of shear-induced cloud-cloud collisions partly counteracts this suppression, leading to a weaker negative or, in some galaxies, positive correlation.

\begin{figure*}
    \centering
    \includegraphics[width=\textwidth]{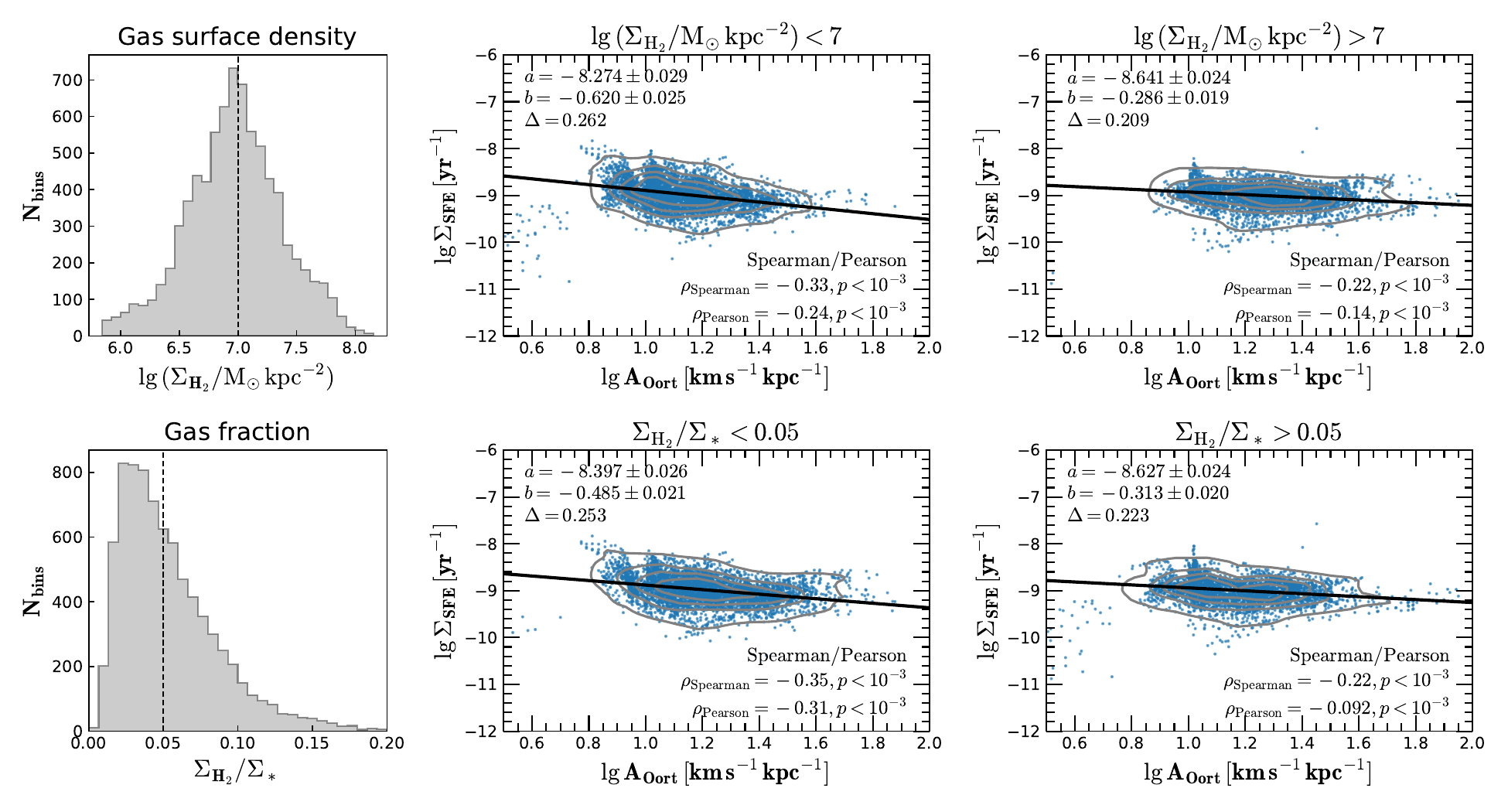}
    \caption{The relation between the star formation efficiency, $\Sigma_{\rm SFE}$, and the galactic shear, $A_{\rm Oort}$, for regions classified as relatively gas-rich and gas-poor. The top row shows the sample split by molecular gas surface density, with gas-poor regions ($\lg\,(\Sigma_{\rm H_2}/{\rm M_{\odot}\,kpc^{-2}})<7$; middle panel) and gas-rich regions ($\lg\,(\Sigma_{\rm H_2}/{\rm M_{\odot}\,kpc^{-2}})>7$; right panel), while the left panel presents the distribution of $\Sigma_{\rm H_2}$ for the full sample, with the adopted division indicated by the dashed line. The bottom row shows the same relation but with the sample divided by molecular gas fraction, $f_{\rm H_2}\equiv\Sigma_{\rm H_2}/\Sigma_{\ast}$, into gas-poor ($f_{\rm H_2}<0.05$) and gas-rich ($f_{\rm H_2}>0.05$) subsamples; the left panel displays the corresponding $f_{\rm H_2}$ distribution. In each $\lg\,\Sigma_{\rm SFE}$–$\lg\,A_{\rm Oort}$ panel, points represent Voronoi bins, contours indicate the kernel density estimate of the data distribution. The coefficients of the best-fitting lines (black solid) $y=a+bx$ and the observed rms scatter $\Delta$ are obtained from the \textsc{lts\_linefit} procedure \citep{Cappellari2013a}. The Spearman rank correlation coefficient, the Pearson correlation coefficient, and their significances are indicated in each panel.}
    \label{fig:SFE_OortA_gas}
\end{figure*}

To investigate whether the regulation of SFE by galactic shear depends on gas density, we divide all Voronoi bins into two groups based on their gas surface density $\Sigma_{\rm H_2}$ or gas fraction $f_{\rm H_2}\equiv \Sigma_{\rm H_2}/\Sigma_{\ast}$. In \autoref{fig:SFE_OortA_gas}, we present the $\Sigma_{\rm SFE}$–$A_{\rm Oort}$ relations for relatively gas-rich ($\lg\,(\Sigma_{\rm H_2}/{\rm M_{\odot}\,kpc^{-2}})>7$ or $f_{\rm H_2}>0.05$) and gas-poor ($\lg\,(\Sigma_{\rm H_2}/{\rm M_{\odot}\,kpc^{-2}})<7$ or $f_{\rm H_2}<0.05$) bins. As expected, the slopes of the best-fitting linear relations for gas-poor bins are steeper (more negative) than those for gas-rich regions, while both the Spearman’s rank correlation coefficients and the Pearson correlation coefficients are less negative for gas-rich bins. As gas density increases, the dependence of SFE on shear weakens due to the increasing influence of cloud–cloud collisions, approaching a balance between two competing mechanisms: enhancement of SFE by cloud–cloud collisions and suppression of SFE by the stabilization of GMC self-gravity. A similar dependence of SFE on gas density was discussed in \citet{Pan2014}, who found in the galaxy IC~342 that the increasing power-law index $N$ of the KS law ($\Sigma_{\rm SFR}\propto\Sigma_{\rm H_2}^{N}$) with higher $\Sigma_{\rm H_2}$ can be interpreted as the influence of cloud–cloud collisions in high-$\Sigma_{\rm H_2}$ regions.

To examine whether the difference between the gas-poor and gas-rich subsamples reflects covariances with other galactic properties, we repeat the analysis after dividing the Voronoi bins by galactocentric radius and stellar mass surface density. Dividing the sample at $R=R_{\rm e}$ yields Spearman coefficients of $-0.29$ and $-0.33$ for the inner and outer regions, respectively, indicating little radial dependence. Dividing by $\Sigma_\ast$ produces a modest change, from $\rho_{\rm Spearman}=-0.32$ at low $\Sigma_\ast$ to $-0.24$ at high $\Sigma_\ast$. Both contrasts are smaller than that obtained by dividing the sample by molecular-gas surface density, for which $\rho_{\rm Spearman}=-0.33$ and $-0.22$ in the gas-poor and gas-rich regions, respectively. Thus, the weaker SFE--$A_{\rm Oort}$ anticorrelation in gas-rich regions cannot be explained solely by a simple radial division, although covariance with stellar mass surface density may contribute to the observed trend.

Although we observe a systematic difference in the $\Sigma_{\rm SFE}$–$A_{\rm Oort}$ relations between the gas-rich and gas-poor subsamples, this difference is not yet statistically significant. This may be due to the relatively narrow range of gas surface densities covered by our sample (see the histograms in \autoref{fig:SFE_OortA_gas}). To further assess the role of gas density, it is essential to extend the comparison to more gas-rich systems, which are more commonly found at higher redshifts \citep[e.g.,][]{Morokuma-Matsui2015,Merida2024}. Owing to the limited sample size, we are also unable to perform a robust statistical analysis of galaxy-to-galaxy variations. A more comprehensive understanding will require a larger and more diverse sample, such as that provided by the Kiloparsec Investigations of Local Objects’ Gas And Star-formation (KILOGAS\footnote{\url{https://kilogas.space/}}) survey, which includes $\sim$500 galaxies spanning a wide range of SFRs, and for which dynamical and stellar population properties are already available through the MaNGA DynPop project.

\subsection{The azimuthal variations of SFE in individual galaxies}\label{subsec:azimuthal_SFE_variation}

\begin{figure*}
    \centering
    \includegraphics[width=0.95\textwidth]{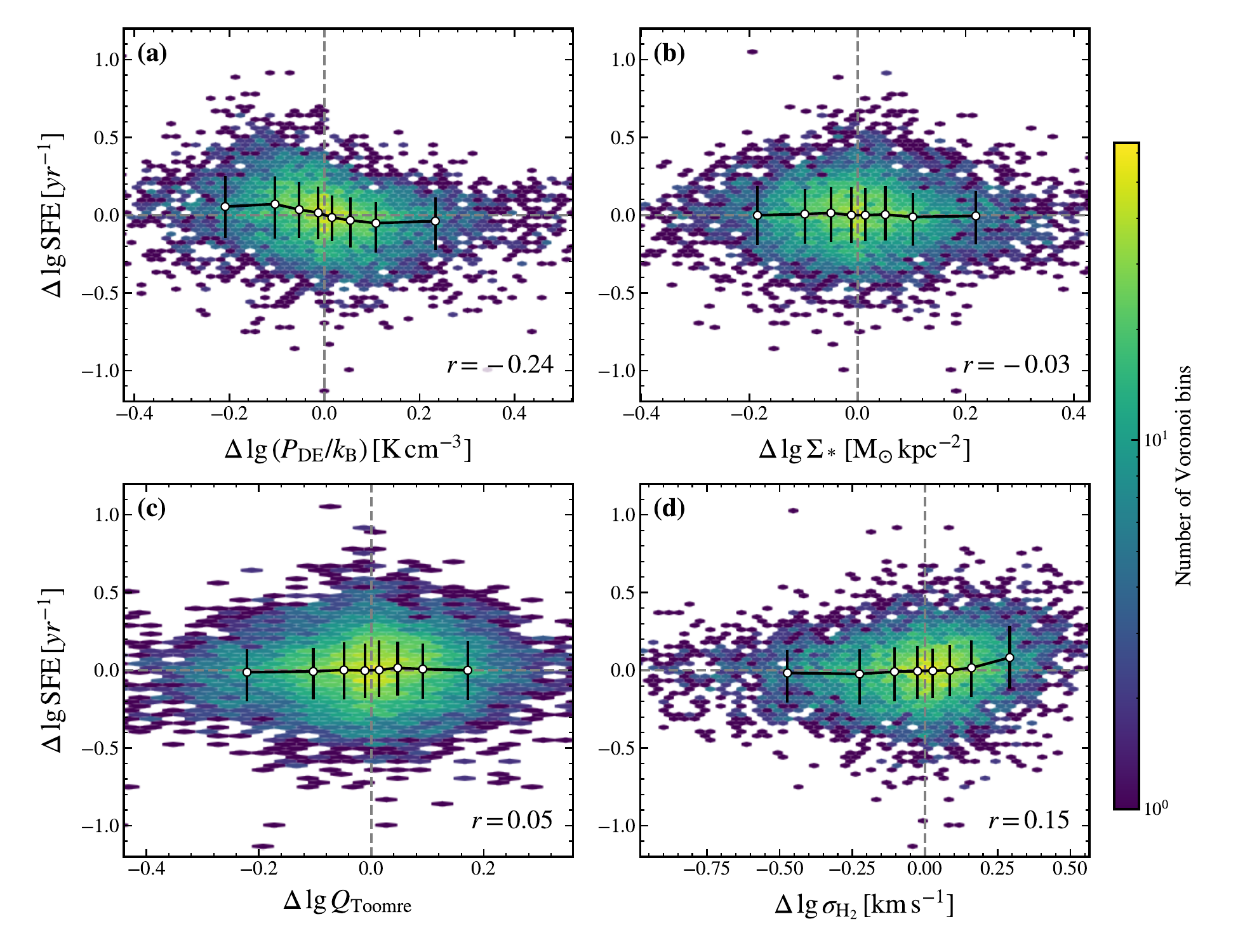}
    \caption{Residuals of $\Sigma_{\rm SFE}$ as a function of the residuals of (a) $P_{\rm DE}$, (b) $\Sigma_{\ast}$, (c) $Q_{\rm Toomre}$, and (d) $\sigma_{\rm H_2}$. For each galaxy, residuals are computed relative to the median value of all Voronoi bins at fixed $A_{\rm Oort}$, thereby removing the dominant radial trends associated with galactic shear. The color bar indicates the number density distribution of all Voronoi bins in the sample, and the median trend associated with 1$\sigma$ error (defined as 16th and 84th percentiles) is also presented. The Pearson correlation coefficient is indicated in each panel.}
    \label{fig:residual_SFE_galprop}
\end{figure*}

In the previous sections, we identify $A_{\rm Oort}$ as the dynamical parameter most strongly associated with variations in $\Sigma_{\rm SFE}$. However, since galactic shear is derived from axisymmetric dynamical models, it primarily traces the radial dependence of SFE. As illustrated in \autoref{fig:maps_example} and \autoref{fig:shear_SFE_individual}, substantial scatter in $\Sigma_{\rm SFE}$ remains at a given $A_{\rm Oort}$ (or equivalently, at a given face-on galactocentric radius), indicating significant azimuthal variations within individual galaxies.

The galactic shear traced by $A_{\rm Oort}$ is determined by the shape of the galactic circular velocity curve and therefore reflects the large-scale dynamics established by the galaxy's gravitational potential. In contrast, quantities such as the stellar mass surface density and mid-plane pressure characterize the local gravitational and ISM environment which may contribute to the azimuthal variations in SFE. To investigate whether these azimuthal variations are linked to other local physical properties, \autoref{fig:residual_SFE_galprop} presents the residuals of $\Sigma_{\rm SFE}$ as a function of the residuals of several non-axisymmetric quantities, including $P_{\rm DE}$, $\Sigma_{\ast}$, $Q_{\rm Toomre}$, and $\sigma_{\rm H_2}$. For each galaxy, residuals are computed relative to the median value of all Voronoi bins at a given $A_{\rm Oort}$, thereby removing the dominant radial trends associated with galactic shear and isolating variations at fixed shear strength.

After accounting for the radial trends associated with galactic shear, the remaining azimuthal variations in $\Sigma_{\rm SFE}$ show only weak correlations with the residuals of the other physical quantities considered here (\autoref{fig:residual_SFE_galprop}). Among them, $P_{\rm DE}$ exhibits the strongest correlation with $\Sigma_{\rm SFE}$ residuals ($r=0.24$), while $\Sigma_{\ast}$ and $Q_{\rm Toomre}$ show virtually no residual dependence. A weak positive correlation is also present for $\sigma_{\rm H_2}$ ($r=0.15$). These results suggest that much of the observed correlation between SFE and several galactic properties reflects their shared radial dependence within galaxies. We note that this analysis does not imply that $A_{\rm Oort}$ is the unique physical driver of SFE, since both galactic shear and other quantities such as $P_{\rm DE}$ and $\Sigma_{\ast}$ ultimately arise from the underlying mass distribution of the galaxy. Rather, it demonstrates that $A_{\rm Oort}$ provides the most effective empirical description of the dominant radial variation in SFE across our sample.

The absence of strong residual correlations indicates that the azimuthal variation in SFE is not fully explained by the axisymmetric quantities considered here. Such variation may arise from unresolved cloud-scale processes, including differences in the evolutionary state of molecular clouds \citep{Chevance2020} and cloud-cloud interactions \citep{Tan2000,Tan2010}. However, $A_{\rm Oort}$ does not capture azimuthally varying shear or non-circular streaming motions induced by bars and spiral structures, which may also regulate cloud collisions and gravitational collapse \citep{Kim2024, Pessa2022}. The residual variations therefore cannot be attributed exclusively to processes below our resolution. Future studies combining dense-gas tracers, higher-resolution observations, and modelling of the full two-dimensional shear field will be required to distinguish cloud-scale processes from non-axisymmetric galactic dynamics.

\section{Conclusions}\label{sec:conlusions}

In this work, we use dust-corrected $\rm H\alpha$ fluxes from the MaNGA DAP datacubes to derive the star formation rate surface density $\Sigma_{\rm SFR}$, and CO intensity maps from the ALMaQUEST survey \citep{Lin2020,Ellison2024} to estimate the molecular gas surface density $\Sigma_{\rm H_2}$. We then define the star formation efficiency as $\Sigma_{\rm SFE}\equiv\Sigma_{\rm SFR}/\Sigma_{\rm H_2}$. By combining these measurements with spatially resolved dynamical properties from the MaNGA DynPop project \citep{Zhu2023,Lu2023}, we investigate the correlations between $\Sigma_{\rm SFE}$ and various local physical properties. After selecting star-forming regions based on the BPT diagram, resulting in $\sim$8000 Voronoi bins from 29 galaxies with reliable dynamical models, we apply a random forest regressor using six local properties (galactic shear $A_{\rm Oort}$, total mass surface density $\Sigma_{\rm total}$, stellar mass surface density $\Sigma_{\ast}$, Toomre stability parameter $Q_{\rm Toomre}$, mid-plane dynamical equilibrium pressure $P_{\rm DE}$, and molecular gas velocity dispersion $\sigma_{\rm H_2}$) to identify the most predictive quantities for $\Sigma_{\rm SFE}$.

The main conclusions are summarized as below.
\begin{enumerate}
    \item In the fiducial analysis, which retains both within-galaxy and galaxy-to-galaxy variations, galactic shear $A_{\rm Oort}$ is the highest-ranked dynamical parameter associated with $\Sigma_{\rm SFE}$ and shows a negative correlation with it ($\rho_{\rm Spearman}=-0.28$). Its predictive importance is higher than those of $\Sigma_{\rm total}$ and $\Sigma_\ast$, although the differences among the highest-ranked parameters are modest (\autoref{fig:RF}). After normalizing each quantity by its galaxy median, $P_{\rm DE}$ becomes the highest-ranked parameter while $A_{\rm Oort}$ remains non-negligible, suggesting that shear is more informative across the full galaxy sample, whereas pressure-related conditions are more closely associated with internal SFE variations.

    \item Among the 18 galaxies with statistically significant ($p<0.05$) $\Sigma_{\rm SFE}-A_{\rm Oort}$ relations, the majority (14 out of 18) show anti-correlations (\autoref{fig:shear_SFE_individual}), suggesting a common suppressive effect of galactic shear ($A_{\rm Oort}$) on $\Sigma_{\rm SFE}$. Nevertheless, the strength of this shear-induced suppression varies substantially from galaxy to galaxy. The presence of four galaxies displaying positive $\Sigma_{\rm SFE}$–$A_{\rm Oort}$ correlations further indicates significant galaxy-to-galaxy variation in the role of shear in regulating SFE.
    
    \item Gas density may play a role in driving the observed galaxy-to-galaxy variation in the SFE-shear relations, as indicated by the weaker anti-correlation observed in more gas-rich regions (\autoref{fig:SFE_OortA_gas}). One possible interpretation is the competition between two mechanisms: the suppressive influence of shear and the background galactic potential, which may stabilize molecular gas or delay its decoupling from the galactic environment \citep{Meidt2025}, and shear-induced cloud-cloud collisions, which become increasingly efficient in high gas-density environments. Further confirmation of this scenario will require a larger and more diverse sample spanning a broader range of gas densities.

    \item While galactic shear is the primary large-scale dynamical parameter regulating SFE, the apparent correlations between SFE and other galactic properties ($P_{\rm DE}$, $\Sigma_{\ast}$, $Q_{\rm Toomre}$, and $\sigma_{\rm H_2}$) are largely driven by their shared radial dependence. Once this radial component is removed, only weak residual correlations remain (\autoref{fig:residual_SFE_galprop}), implying that azimuthal variations in SFE are likely driven by local cloud-scale processes below our resolution scale.
\end{enumerate}

\begin{acknowledgments}

We thank Yung-Chau Su for providing the CO-based rotation curve measurements. We thank the anonymous referee for constructive comments that improved the clarity and robustness of this work. We also thank Luis Ho, Martin Bureau, and Junfeng Wang for helpful discussions. This work is partly supported by the National Science Foundation of China (grant No. 11821303 to S.M.). KZ acknowledges support from the China Postdoctoral Science Foundation (No. 2026M793514). HAP acknowledges support from the National Science and Technology Council of Taiwan under grant 113-2112-M-032-014-MY3.

This paper makes use of the
following ALMA data:
ADS/JAO.ALMA\#2015.1.01225.S,\\
ADS/JAO.ALMA\#2017.1.01093.S,\\
ADS/JAO.ALMA\#2018.1.00558.S,\\
ADS/JAO.ALMA\#2018.1.00541.S.
ALMA is a partnership
of ESO (representing its member states), NSF (USA), and
NINS (Japan), together with NRC (Canada), MOST and ASIAA(Taiwan), and KASI (Republic of Korea), in cooperation with the Republic of Chile. The Joint ALMA Observatory is operated by ESO, AUI/NRAO, and NAOJ. The National Radio Astronomy Observatory is a facility of the National Science Foundation operated under cooperative agreement by Associated Universities, Inc.

Funding for the Sloan Digital Sky Survey IV has been provided by the Alfred P. Sloan Foundation, the U.S. Department of Energy Office of Science, and the Participating Institutions. 

SDSS-IV acknowledges support and resources from the Center for High Performance Computing at the University of Utah. The SDSS website is www.sdss.org.

SDSS-IV is managed by the Astrophysical Research Consortium for the Participating Institutions of the SDSS Collaboration including the Brazilian Participation Group, the Carnegie Institution for Science, Carnegie Mellon University, Center for Astrophysics | Harvard \& Smithsonian, the Chilean Participation Group, the French Participation Group, Instituto de Astrof\'isica de Canarias, The Johns Hopkins University, Kavli Institute for the Physics and Mathematics of the Universe (IPMU) / University of Tokyo, the Korean Participation Group, Lawrence Berkeley National Laboratory, Leibniz Institut f\"ur Astrophysik Potsdam (AIP),  Max-Planck-Institut f\"ur Astronomie (MPIA Heidelberg), Max-Planck-Institut f\"ur Astrophysik (MPA Garching), Max-Planck-Institut f\"ur Extraterrestrische Physik (MPE), National Astronomical Observatories of China, New Mexico State University, New York University, University of Notre Dame, Observat\'ario Nacional / MCTI, The Ohio State University, Pennsylvania State University, Shanghai Astronomical Observatory, United Kingdom Participation Group, Universidad Nacional Aut\'onoma de M\'exico, University of Arizona, University of Colorado Boulder, University of Oxford, University of Portsmouth, University of Utah, University of Virginia, University of Washington, University of Wisconsin, Vanderbilt University, and Yale University.
\end{acknowledgments}

\appendix

\section{Random forest regression for azimuthally averaged quantities}
\label{appendix:rf_azimuthal}

The Oort parameter $A_{\rm Oort}$ adopted in our resolved analysis is derived from an axisymmetric circular velocity curve and is therefore a function of galactocentric radius only, whereas most of the other quantities, including the SFE, are measured from two-dimensional maps. To place all quantities on the same dimensional footing, we repeat the random forest analysis using azimuthally averaged measurements. For each galaxy, the valid Voronoi bins are grouped into radial annuli of width $0.2 R_{\rm e}$. We retain only annuli containing at least three Voronoi bins. Each Voronoi bin is weighted by its number of constituent spaxels, such that the resulting averages approximate area-weighted annular quantities. All positive quantities are averaged in linear space before being transformed to logarithmic units. The molecular gas velocity dispersion is instead weighted by molecular-gas mass, with weights proportional to $N_{\rm spaxel}\Sigma_{\rm H_2}$. The annular SFE is defined as the area-weighted mean of the local SFE measurements. Consequently, each galaxy annulus contributes only one data point to the regression, avoiding the repeated use of the same radial $A_{\rm Oort}(R)$ value for multiple azimuthal Voronoi bins.

\begin{figure*}
    \centering
    \includegraphics[width=\linewidth]{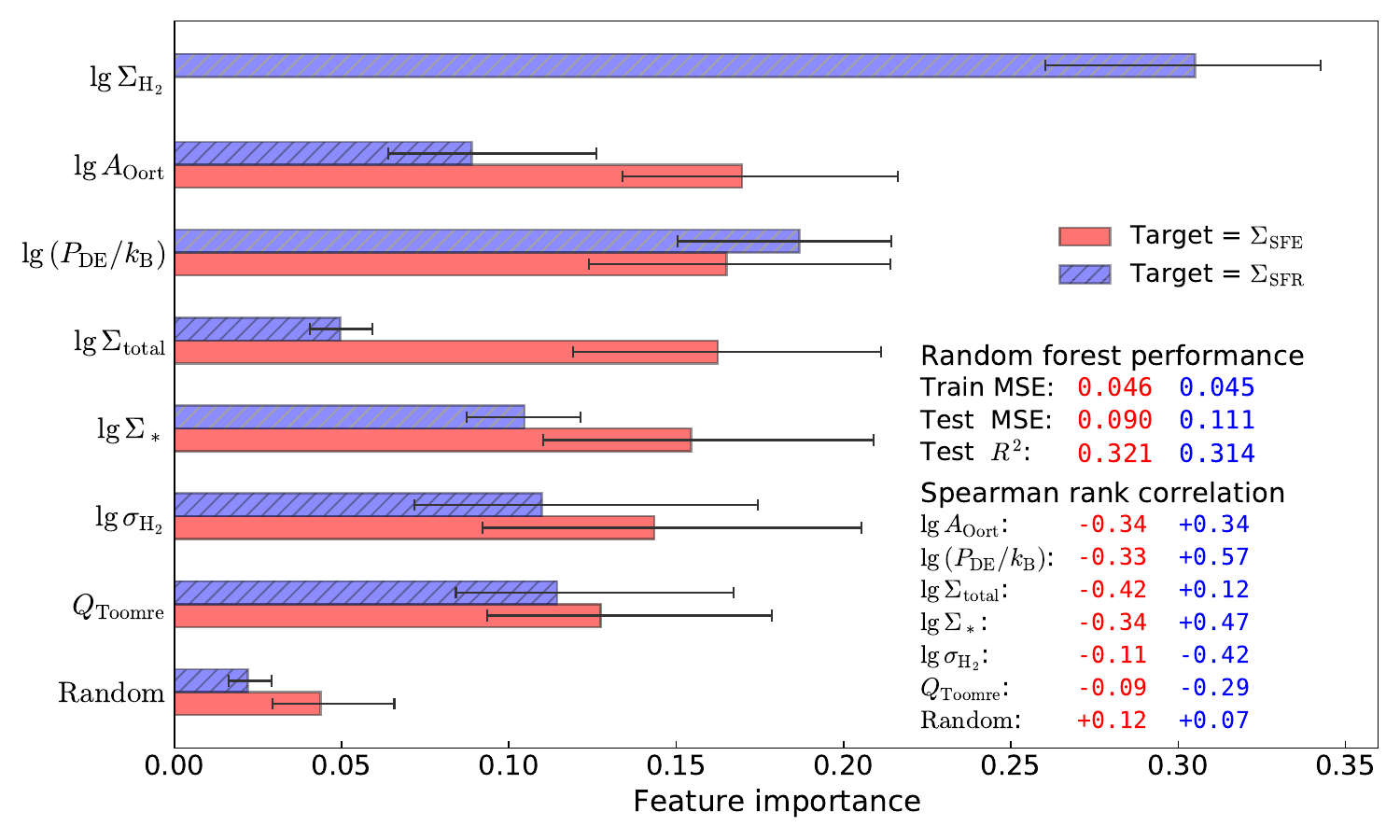}
    \caption{Similar to \autoref{fig:RF}, but showing the random forest feature importances obtained using azimuthally averaged quantities in radial annuli of width $0.2R_{\rm e}$. Red bars show the results with the area-weighted SFE as the regression target, while blue hatched bars show the corresponding results for the area-weighted SFR. Bar lengths indicate the median feature importance obtained from 300 galaxy-level bootstrap realizations, and the error bars enclose the corresponding 16th-84th percentile intervals. In each bootstrap realization, galaxies are resampled with replacement, and all radial annuli belonging to each selected galaxy are retained.}
    \label{fig:RF_azimuthal}
\end{figure*}

We train separate random forest regressors for the azimuthally averaged SFE and SFR, following the same analysis framework as that used for
\autoref{fig:RF}. The model hyperparameters are reoptimized for the azimuthally averaged data set. Uncertainties in the feature importances are estimated using 300 galaxy-level bootstrap realizations. In each realization, galaxies are resampled with replacement, and all radial annuli belonging to each selected galaxy are retained. \autoref{fig:RF_azimuthal} shows the median feature
importances and the corresponding 16th-84th percentile intervals. The
qualitative ranking of $A_{\rm Oort}$ in predicting the SFE remains consistent with that obtained from the resolved Voronoi bin analysis, although the median importance of $A_{\rm Oort}$ is lower in the annular analysis. We therefore conclude that our main result is not driven solely by the comparison of a one-dimensional shear measurement with two-dimensional resolved quantities.

\section{Comparison between the JAM-based circular velocity curves and CO rotation curves}
\label{appendix:cvc_comparison}

\begin{figure}
    \centering
    \includegraphics[width=\textwidth]{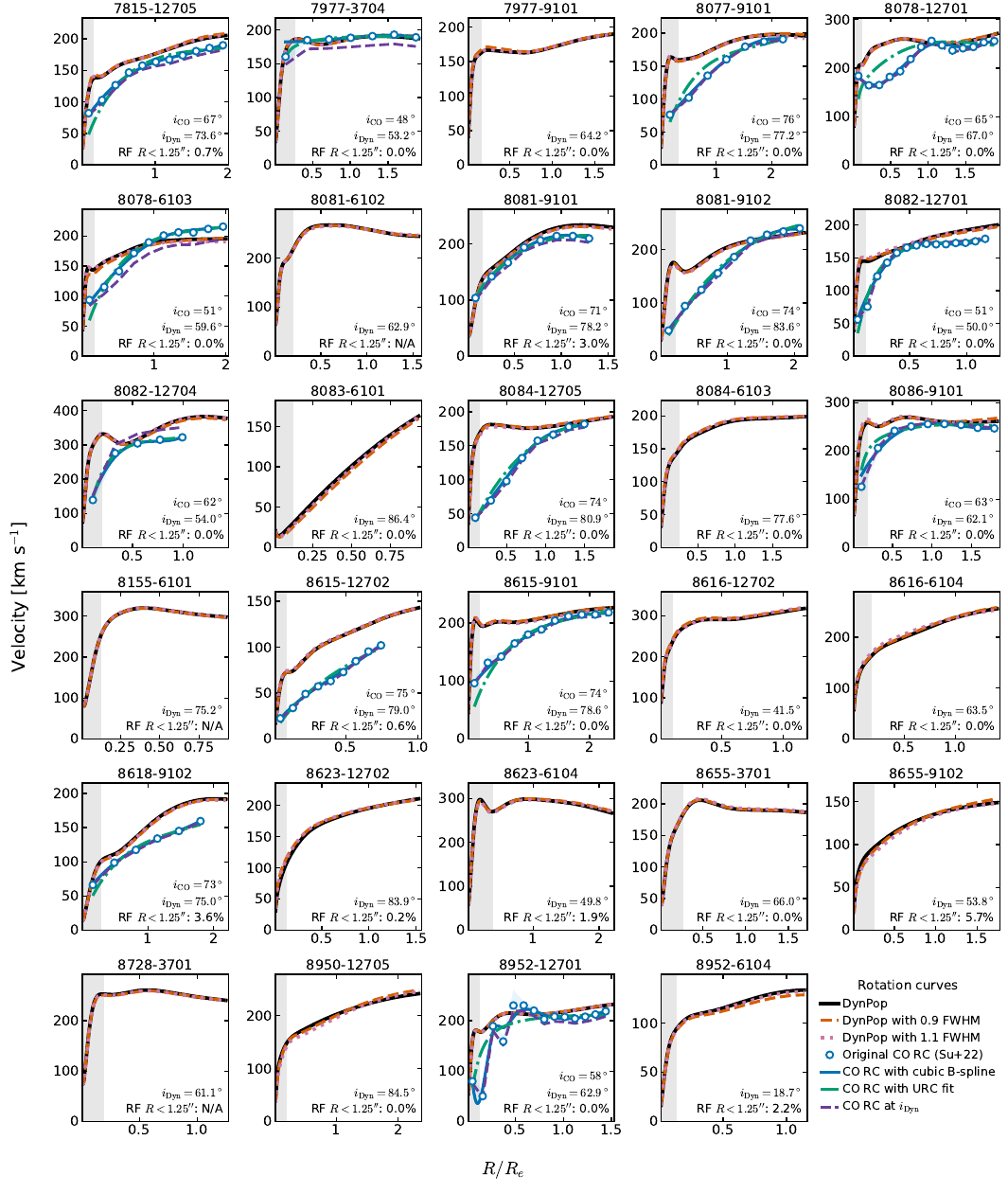}
    \caption{ Comparison of the JAM-based circular-velocity curves and CO rotation curves for the 29 galaxies. Black solid, orange dashed, and magenta dotted curves show the JAM results obtained using the nominal, $0.9\times$, and $1.1\times$ MaNGA PSF FWHM, respectively. For the 15 galaxies overlapping with \citet{Su2022}, open blue circles and light-blue bands denote the original \textsc{3DBarolo} CO measurements and their uncertainties. Blue solid and green dash-dotted curves show their regularized cubic B-spline and URC representations, respectively. Purple dashed curves show the CO velocities rescaled from $i_{\rm CO}$ to $i_{\rm Dyn}$ using \autoref{eq:CO_RC_inc}. Both inclinations are listed in the lower-right corner of the relevant panels. The gray region marks $R<1.25''$, half the adopted $2.5''$ MaNGA PSF and ALMA CO beam FWHM. The fraction of Voronoi bins within this gray region relative to all Voronoi bins used in the random forest analysis is also reported in each panel.}
\label{fig:RC_DynPop_psf_CO}
\end{figure}

Figure~\ref{fig:RC_DynPop_psf_CO} compares the JAM-based circular velocity curves (from the DynPop catalog)
with the CO rotation curves available for the selected galaxy sample. This comparison is intended both to quantify the sensitivity of the JAM results to the adopted MaNGA PSF and to examine the consistency between the JAM-based and CO-based rotation curves.

The spatial variation of MaNGA PSF across an individual data cube is less than 10 $\%$ level \citep{Law2016}. We therefore repeat the JAM modelling after multiplying the measured PSF FWHM by factors of 0.9 and 1.1, while all other modelling assumptions are kept unchanged. Although some differences are visible in the central regions of individual galaxies, the three circular velocity curves with different PSF FWHM generally remain similar over the radial ranges sampled by the Voronoi bins in our resolved
analysis. Thus, a plausible 10 $\%$ variation in the adopted MaNGA PSF does not qualitatively change the inferred circular velocity curves or our conclusions based on $A_{\rm Oort}$.

For the 15 galaxies overlapping the sample of \citet{Su2022}, we also show the original CO rotation curve measurements \citep[][Figure~C1]{Su2022} obtained by fitting the CO data cubes with \textsc{3DBarolo} \citep{DiTeodoro2015}. To suppress measurement noise of $A_{\rm Oort}$ due to direct differentiation of the discrete CO measurements, we construct two smoothed CO rotation curves. The first is a regularized cubic B-spline, which provides a flexible description of the observed radial structure while suppressing small-scale
fluctuations. The second is a parametric universal rotation curve \citep[URC;][]{Persic1996} fit, which imposes a more restrictive and globally smooth functional form. They provide two alternative CO-based estimates of $A_{\rm Oort}$ and allow us to assess the sensitivity of the random forest results to the adopted $A_{\rm Oort}$ measurements.

The CO and JAM-based analyses do not necessarily adopt the same disk inclination, part of the difference between the two rotation curves may arise from their different
inclination estimates. As a diagnostic, we rescale the original CO rotation velocities to the JAM-inferred inclination using
\begin{equation}\label{eq:CO_RC_inc}
V_{\rm CO}(i_{\rm Dyn})
 =
V_{\rm CO}(i_{\rm CO})
\frac{\sin i_{\rm CO}}{\sin i_{\rm Dyn}}.
\end{equation}
The inclinations $i_{\rm CO}$ and $i_{\rm Dyn}$, which are derived from \textsc{3DBarolo} and JAM, are listed in the lower-right corner of the corresponding
panels. Rescaling the CO velocities to $i_{\rm Dyn}$ does not generally remove the differences between the CO-based and JAM-based rotation curves, indicating that the adopted inclination alone cannot account for these differences. 

The gray region in each panel marks $R<1.25''$, corresponding to half of the approximately $2.5''$ MaNGA PSF FWHM and the $2.5''$ ALMA CO beam FWHM used for the ALMaQUEST data. Measurements within this region are expected to be the most sensitive to finite spatial resolution and beam-smearing effects. For each galaxy, \autoref{fig:RC_DynPop_psf_CO} also reports the fraction of Voronoi bins within $R<1.25''$ relative to the total number of Voronoi bins used in the random forest analysis. These fractions are small for most galaxies, indicating that our random forest results are not dominated by measurements from the central resolution-affected region.

\bibliography{ref}{}
\bibliographystyle{aasjournalv7}

\end{document}